\documentclass[aps,prx,amsmath,twocolumn,amssymb,superscriptaddress,longbibliography]{revtex4-2}
\usepackage[T1]{fontenc}
\usepackage[colorlinks=true,citecolor=blue,linkcolor=magenta]{hyperref}
\usepackage{amssymb,amsfonts,amsmath}
\usepackage{color}
\usepackage[apple mac]{inputenc}
\usepackage[english]{babel}
\usepackage{times}
\usepackage{latexsym}
\usepackage{verbatim}
\usepackage{graphicx}
\usepackage{epsf}
\usepackage{subfigure}
\usepackage{bm}
\usepackage{amssymb}
\usepackage{stackrel}
\usepackage{float}

\usepackage{ulem} 

\graphicspath{{./},{figures/},}

\usepackage{epstopdf}

\definecolor{Nathanblue}{rgb}{0.,0.24,0.51}
\newcommand{\blue}{\color{Nathanblue}}
\definecolor{orange}{rgb}{0.76,0.44,0.0}

\usepackage{color}

\usepackage{mathtools}

\def\be{\begin{equation}}
\def\ee{\end{equation}}

\begin{document}


\title{{\blue Controlled Buildup of Half-Quantized Thermal Conductance \\ in an Engineered Chiral Spin Liquid Platform}}

\author{Bo-Ye Sun}
\email[]{boyesun@ytu.edu.cn}
\affiliation{YanTai University, Yantai, Shandong, 264005, People's Republic of China}

\author{Baptiste Bermond}
\affiliation{Laboratoire Kastler Brossel, Coll\`ege de France, CNRS, ENS-Universit\'e PSL, Sorbonne Universit\'e, 11 Place Marcelin Berthelot, 75005 Paris, France}

\author{Lucila Peralta Gavensky}
\affiliation{Center for Nonlinear Phenomena and Complex Systems, Universit\'e Libre de Bruxelles, CP 231, Campus Plaine, B-1050 Brussels, Belgium}
\affiliation{International Solvay Institutes, 1050 Brussels, Belgium}

\author{Marin Bukov}
\affiliation{Max Planck Institute for the Physics of Complex Systems, N\"othnitzer Str.~38, 01187 Dresden, Germany}

\author{Zheng-Wei Zhou}
\affiliation{Anhui Province Key Laboratory of Quantum Network,
University of Science and Technology of China, Hefei, 230026, China}
\author{Nathan Goldman}
\affiliation{Laboratoire Kastler Brossel, Coll\`ege de France, CNRS, ENS-Universit\'e PSL, Sorbonne Universit\'e, 11 Place Marcelin Berthelot, 75005 Paris, France}
\affiliation{Center for Nonlinear Phenomena and Complex Systems, Universit\'e Libre de Bruxelles, CP 231, Campus Plaine, B-1050 Brussels, Belgium}
\affiliation{International Solvay Institutes, 1050 Brussels, Belgium}

\begin{abstract}
We study thermal transport along the edge of a small chiral-spin-liquid device coupled to two Ising-chain reservoirs, a platform suitable for quantum-engineered systems. Adiabatically switching on the tunnel couplings to the reservoirs generates a thermal current that dynamically builds up and reaches a quasi-steady-state regime. In this time window, the two-terminal thermal conductance can approach half-quantized values -- a hallmark of Majorana-mediated transport -- under finely tuned conditions. The results agree with a steady-state Landauer-B\"uttiker description for sufficiently large reservoirs, where energy-resolved transmission rates help identify the optimal parameters to achieve the half-quantized conductance. This work provides a controllable platform to investigate topological thermal transport in engineered spin systems, such as realized in cold-atom and Rydberg-atom settings.
\end{abstract}
\date{\today}

\maketitle

\section{\label{sec:intro}Introduction}
The Kitaev honeycomb model~\cite{kitaev2006anyons} has emerged as a cornerstone in the study of quantum spin liquids, providing a rich platform for exploring exotic physics and topological phenomena~\cite{semeghini2021probing,satzinger2021realizing,xu2022digital,zhou2022probing}, and motivating the search for, and study of, so-called Kitaev materials~\cite{trebst2022kitaev,hermanns2018physics,knolle2019field,takagi2019concept}. Known for its exact solvability, the model offers a rigorous theoretical understanding of its ground state and corresponding phases~\cite{kitaev2006anyons,sun2023engineering}. It exhibits a gapped, Abelian spin-liquid phase whenever one bond coupling dominates over the other two, and hosts a gapless spin liquid whenever all three couplings are of comparable strength (so that none exceeds the sum of the other two). 
Particularly for the gapless phase, with an external magnetic field breaking time-reversal symmetry and opening up a topological Majorana gap, a chiral spin liquid appears, which holds profound implications for both fundamental research and quantum computation~\cite{kitaev2003fault,stern2013topological,andersen2022observation,lensky2022graph,harle2023observing}. Extensive experimental and theoretical efforts have been dedicated to identifying clear signatures of the quantum spin liquid phase in realistic systems~\cite{knolle2014dynamics,knolle2015dynamics,schmitt2015dynamical,banerjee2017neutron,smith2016majorana,chen2018nonabelian,feldmeier2020local,mizoguchi2020oriented}. Among these signatures, the half-quantized thermal conductance has been recognized as a hallmark of the chiral spin liquid phase. Unlike the quantized Hall effect in Chern insulators, this phenomenon features neutral Majorana fermions as carriers instead of electrons, offering a direct probe of its unconventional thermal behavior~\cite{nasu2017thermal,joy2022dynamics}.

Experimental observations of half-quantized thermal conductance in $\alpha$-RuCl$_3$~\cite{kasahara2018majorana,bruin2022robustness,yamashita2020sample}, a chiral spin liquid candidate, mark a significant breakthrough in solid-state systems. However, such systems inherently involve complexities, including phonon scattering and material imperfections, which can obscure the intrinsic physics of the Kitaev model. In contrast, cold atom systems offer an alternative approach, with their highly controlled and tunable settings enabling a cleaner and more direct simulation of the model~\cite{goldman2016topological,cooper2019topological,leonard2023realization,de2019observation,verresen2021prediction}. Building on the theoretical and experimental attempts for realizing chiral spin liquids in cold atoms~\cite{sun2023engineering,chen2024realization,kalinowski2023non-abelian,evered2025probing,will2025probing}, this work is motivated by the need to establish methods for observing half-quantized thermal conductance in such platforms, which hold the potential to advance our understanding of quantum spin liquids.

\begin{figure}[]
    \centering
    \includegraphics[width=0.48\textwidth]{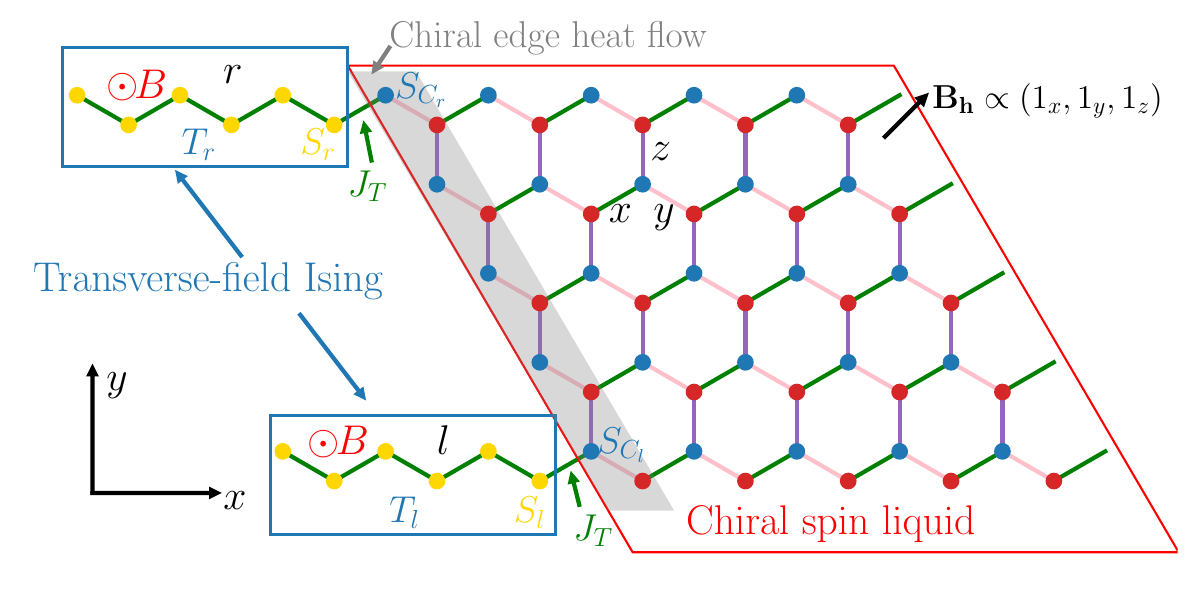}
    \caption{\textbf{Schematic of the proposed system.} The chiral spin liquid (red and blue sites) is coupled to two reservoirs (yellow sites) along the same edge. Each reservoir is modeled as a one-dimensional transverse-field Ising chain with open boundary conditions, and $\odot$ denotes the local transverse fields acting on the Ising spins. The green ($x$), pink ($y$), and purple ($z$) bonds indicate the direction $\alpha$ of the bonds $\langle i,j\rangle_\alpha$ in the Hamiltonian in Eq.~\eqref{eq:Hc}. The label $l/r$ in the figure denotes the left/right reservoirs, respectively, $S_{l/r}$ as well as $S_{C_{l/r}}$ denotes the coupling points between the reservoirs and the channel in Eq.~\eqref{eq:Ht}, ${\bf B_h}$ represents the magnetic field that induce $h$ in Eq.~\eqref{eq:Hc}, and $J_T$ denotes the tunnel coupling between each reservoir and the Kitaev channel. A temperature bias ($T_l \ne T_r$) drives heat transport through the chiral edge channel, highlighted by the gray shaded region.    
    }
    \label{Fig1}
\end{figure}

Inspired by the proposals of Refs.~\cite{sun2023engineering,kalinowski2023non-abelian,chen2024realization} for realizing a chiral spin liquid in ultracold atoms, we propose a two-terminal heat-transport setup suitable for cold-atom quantum simulators. The reservoirs are modeled as finite transverse-field Ising chains, each coupled to a central Kitaev spin-liquid channel~\cite{wu2018crossovers,borish2020transverse}, as shown in Fig.~\ref{Fig1}. The transverse-field Ising model shares similarities with the Kitaev model in that both are spin-interaction-based systems. This resemblance allows for the same cold atom system to be employed in constructing the setup, thereby simplifying its experimental implementation. Furthermore, the transverse-field Ising model becomes gapless at its critical point~\cite{pfeuty1970the} (see App.~\ref{app:JW_spectrum}), making it particularly suitable for studying low-temperature transport: the vanishing excitation gap guarantees low-energy modes that remain thermally populated even as the temperature tends to zero. Specifically, it provides an ideal platform to investigate the half-quantized thermal conductance arising from the transport of Majorana fermions along the edge channel of the Kitaev model. 

With this configuration, we demonstrate that even with finite reservoirs, a stable heat flow region can be achieved, and the corresponding thermal conductance can approach a half-quantized value under very low temperatures. 
Using the Landauer-B\"uttiker formalism~\cite{haug2008quantum,datta1995electronic}, we offer a comprehensive explanation of the observed thermal conductance behavior. This provides experimentalists with concrete guidelines for detecting half-quantized thermal conductance in cold atom setups.

In Sec.~\ref{sec:Model}, we introduce the system setup and Hamiltonian, present the time-dependent build-up of energy current leading to the thermal conductance, and outline the Landauer-B\"uttiker approach used for interpretation. In Sec.~\ref{sec:measure}, we demonstrate that the Landauer-B\"uttiker formalism is consistent with direct numerical simulation results when the reservoirs are sufficiently large, and then use this formalism to investigate how the coupling strength and temperature influence the system. In Sec.~\ref{sec:num_thermal}, we calculate the thermal conductance and show its dependence on the coupling strength and the temperature. Finally, we summarize our findings in Sec.~\ref{sec:sum}.

\section{\label{sec:Model} Model and Methods}

In this section, we first introduce the Hamiltonian of the model, and then describe a realistic protocol leading to the desired system configuration; this includes the preparation of the initial state, the connection of the reservoirs to the central channel, and the observables relevant for probing the system's thermal conductance. Finally, we outline the computational methods employed in our simulations, corresponding to each step of this protocol.

\subsection{The engineered system and reservoirs}
Our setting involves a Kitaev honeycomb model connected to two Ising-chain reservoirs; see Fig.~\ref{Fig1}. The theoretical framework described below assumes that the Kitaev system is static, which could be realized within a Rydberg-atom platform~\cite {chen2024realization}. Generalizations of our framework to Floquet-engineered Kitaev settings~\cite{sun2023engineering,kalinowski2023non-abelian} could be envisaged along the lines of Refs.~\cite{Kundu_2013,Farrell2015,Yap_2017,Salerno_2019}.

The Hamiltonian of our system-reservoirs setting is given by $H=H_C+H^l_R+H^r_R+H_T$, with
\begin{eqnarray}
H_C&=&-J_0\sum_{\langle i,j \rangle_ \alpha}S^\alpha_iS^\alpha_j+h\sum_{[ijk]_{\alpha\beta\gamma}}S^\alpha_iS^\beta_jS^\gamma_k, \label{eq:Hc}\\
H_R^{l/r}&=&\sum_{j\in l/r} -J_R S^x_jS_{j+1}^x+BS^z_j,\\
H_T&=&-J_TS^x_lS^x_{C_l}-J_TS^x_rS^x_{C_r}.\label{eq:Ht}
\end{eqnarray}
Here, we work in natural units and set the reduced Planck constant 
$\hbar=1$, the Land\'e $g$-factor $g=1$, and the Bohr magneton $\mu_B=1$. $H_C$ denotes the chiral spin liquid channel~\cite{kitaev2006anyons}, where $J_0$ denotes the strength of the spin coupling, and $h$ specifies the magnitude of the external ``magnetic" field responsible for the topological gap. The spin-$1/2$ operators $S_i^\alpha$ at site $i$ along the direction $\alpha$ satisfy the standard spin commutation relations $[S^\alpha_i,S^\beta_j]{=}i\delta_{ij}\varepsilon^{\alpha\beta\gamma}S^\gamma_j$. In the Kitaev model, three distinct bond types connecting nearest-neighbor sites define the direction $\alpha=x$, $y$, and $z$ of the involved spin-spin interaction. Their corresponding configurations are shown by the green, pink, and purple bonds in Fig.~\ref{Fig1}. In the Hamiltonian, we denote these bonds using $\langle i,j\rangle_{\alpha}$, which means that for any two neighboring sites $i$ and $j$, the corresponding $\alpha$ is determined according to the configurations shown in the figure. Furthermore, the notation $[ijk]_{\alpha\beta\gamma}=\langle i,j\rangle_{\alpha}
\langle j,k\rangle_{\gamma}$ in the Hamiltonian represents that given any three neighboring sites $i$, $j$, and $k$, the corresponding indices $\alpha$ and $\gamma$ are determined by the bonds $\langle i,j\rangle_{\alpha}$ and $\langle j,k\rangle_{\gamma}$ shown in Fig.~\ref{Fig1}, where all indices are different, i.e., $\beta \neq \alpha \neq \gamma$.

The term $H_R^{l/r}$ describes the transverse-field Ising model representing the two reservoirs, labeled as $l$ (left) and $r$ (right). Here, the magnetic field strength $B$ is set to half the coupling strength $J_R$ to ensure a gapless excitation spectrum (see App.~\ref{app:JW_spectrum}), and $J_R$ is set equal to $J_0$ for simplicity. 
The final term $H_T$ represents the coupling between the reservoirs and the Kitaev channel, with strength $J_T$. In particular, $S_{l/r}$ denotes the spins in the reservoirs located at the points that connect to the channel, while $S_{C_{l/r}}$ represents the spins in the channel connected to these reservoir points, as shown in Fig.~\ref{Fig1}. 

\subsection{The protocol: preparation and measurement\label{sec:protocol}}

Having introduced the Hamiltonian, we now describe the protocol used to configure the system in a cold-atom implementation, along with the observables relevant to thermal-conductance measurements. 
To accurately measure the thermal conductance, the reservoir temperatures must be known. Therefore, the reservoirs are first decoupled from the Kitaev channel, allowing both the reservoirs and the channel to reach thermal equilibrium independently. This procedure defines well-characterized temperatures for each subsystem. Afterward, the reservoirs are reconnected to the channel to enable further evolution of the system. To limit high-energy excitations, a gradual ramping approach is applied during the reconnecting process:
\begin{equation}
H_T(t)=-\frac{J_T}{1+e^{-(t-\sigma t_s)/t_s}}[S^x_l(t)S^x_{C_l}(t)+S^x_r(t)S^x_{C_r}(t)],\label{eq:recon}
\end{equation}
where $t_s$ sets the characteristic time scale of the ramp (controlling its steepness), and $\sigma t_s$ sets the center time of the ramp (i.e., a time offset).  While a small $t_s$ may lead to undesirable high-energy excitations, a very large $t_s$ would result in long ramp durations, which are experimentally challenging due to finite system lifetimes. In our simulations, we choose $t_s = 100J_0$ to balance these considerations. Here, $S^\alpha_{i}(t)$ denotes the corresponding spin operators at time $t$. These time-dependent operators are defined in the Heisenberg picture:
\begin{equation}
    S^\alpha_{i}(t)=e^{iHt} S^\alpha_{i} e^{-iHt}.\label{eq:spin_evo}
\end{equation}
We assume that the reservoirs are predominantly isolated from external energy exchanges, apart from their interactions with the Kitaev system, both during the reconnection process and throughout the subsequent time evolution. This means that the reservoirs are not required to maintain a fixed temperature during this period; instead, the evolution of their quasiparticle excitations is entirely governed by the given Hamiltonian in Eqs.~\eqref{eq:Hc} to \eqref{eq:Ht}.

With these definitions of the Hamiltonians, the energy variation of the reservoirs, i.e., the energy  current flowing into the reservoirs, can be obtained by using Ehrenfest's theorem:
\begin{eqnarray}    
    J^{l/r}_E(t)&=&\langle \dot{H}^{l/r}_R(t)\rangle=i\langle[H(t),H^{l/r}_R(t)] \rangle \nonumber\\
    &=&-J_T B \langle S^y_{l/r}(t) S_{C_{l/r}}^x(t) \rangle.\label{eq:JE}
\end{eqnarray}
Here, $\langle \cdot \rangle$ denotes the quantum expectation value, i.e. 
$\langle O(t)\rangle={\rm Tr}[\rho(t)O(t)]$ with $\rho(t)$ the density matrix of the system. Since the elementary excitations in both the reservoirs and the channel are quasiparticles with effectively zero chemical potential at all times, this energy current is just the thermal current. This means that the time evolution of the thermal current through each reservoir can be directly obtained by measuring the (equal-time) spin correlation function $\langle S^y_{l/r}(t) S_{C_{l/r}}^x(t) \rangle$ between the corresponding reservoir and the channel.

\subsection{Computational methods}

The Kitaev honeycomb model can be solved exactly. In our system, this property is made explicit through the use of the Jordan-Wigner transformation, enabling a detailed analysis of the system's dynamics~\cite{feng2007topological,chen2007exact,chen2008exact,sun2023engineering}. In particular, the spin-pair terms $S^y_{l/r}S^x_{C_{l/r}}$ can be mapped to Majorana fermions via the same transformation, allowing the time evolution of the thermal current Eq.~\eqref{eq:JE} to be directly computed. In this subsection, we provide a general description of the numerical method used to perform this direct time-evolution calculation, with more details given in App.~\ref{app:thermal_current}. 

In addition to the direct numerical simulation approach, we also introduce the frequency-dependent transmission rate derived from the Landauer-B\"uttiker formalism. This method provides an independent and complementary perspective, allowing us to gain further insights into the transport properties of the system and to cross-check the numerical findings. Our analysis is restricted to the vortex-free sector of the chiral spin liquid channel~\cite{sun2023engineering}, under the assumption of very low temperature. This regime is expected to be valid when the temperature is smaller than $0.01J_0$, as shown by the quantum Monte Carlo study of Nasu et al.~\cite{nasu2017thermal}.

\subsubsection{\label{sec:exact} Direct time-evolution simulation}
As described in Sec.~\ref{sec:protocol}, the reservoirs are initially decoupled from the central channel, and each is prepared in thermal equilibrium at a known temperature. Under this condition, the excitations in both the reservoirs and the channel follow the Fermi distribution. This allows us to construct the initial density matrix of the full system and transform it into the Majorana representation. The resulting density matrix takes the form
\begin{equation}
    \rho_{i,j}=2\sum_{\xi_k>0}[(U^\dagger_{k i}U_{jk}-U_{ik} U^\dagger_{k j})n_F(\xi_k,T)+U_{ik}U^\dagger_{kj}],
    \label{eq:rho_used}
\end{equation}
where $n_F(\xi_k,T)$ is the Fermi distribution function at the excitation energy $\xi_k$ and temperature $T$, and $U_{ik}$ is the transformation matrix relating Majorana fermions to the system's quasi-particle excitations (see App.~\ref{app:thermal_current} for details). Notably, this transformation introduces a term independent of excitation distributions -- the last term in Eq.~\eqref{eq:rho_used}. This term corresponds to the density matrix at zero temperature and gives the ground state energy. It also contributes to the thermal conductance, and this contribution is referred to as the influence of the ground state 
in this work. With the initial state and the exactly solvable Hamiltonian, the time evolution of the thermal current $J^{l/r}_E(t)$ can be obtained directly via the Heisenberg equation of motion [see Eq.~\eqref{eq:time_evo}]. In our simulations, we compute the full dynamics, including the reconnection process [see Eq.~\eqref{eq:recon}].

Having obtained the energy current $J^{l/r}_E(t)$ flowing into each reservoir, we now turn to the thermal conductance. For a two-terminal system, thermal conductance is conventionally defined once a steady-state heat flow is established, where the magnitudes of the currents at the two reservoirs become equal and opposite. At that point, the conductance can be extracted from the thermal current of either reservoir using the standard definition~\cite{nasu2017thermal,sumiyoshi2013quantum}. To monitor the approach toward this steady state, we introduce a time-resolved quasi-thermal conductance for each reservoir:
\begin{equation}
\kappa_{l/r}(t)/T = 2J^{l/r}_E(t) / |T_l^2 - T_r^2|. \label{eq:kappalr}
\end{equation}
This expression is based on the definition $\kappa_{l/r} = J^{l/r}_E /|T_l - T_r|$, i.e., the corresponding thermal current divided by the temperature difference, and the average temperature is taken as $T=(T_l+T_r)/2$.  This quasi-thermal conductance shares the same physical units as the conventional thermal conductance and offers a practical means to monitor the onset of stable transport, even in the transient regime.

In our study, the quasi-thermal conductance is computed via direct time evolution simulation, where the system is initially isolated and then evolves out of equilibrium as the reservoirs are dynamically coupled to the Kitaev channel. This time-evolution approach explicitly captures the transient dynamics of the system, in contrast to previous calculations based on the Kubo formula~\cite{nasu2017thermal} and the Landauer-B\"uttiker formalism~\cite{wan2024quarter,yan2021a,huang2018disorder,arrachea2009thermal}, both of which assume a steady-state regime and thus only apply once time-translation invariance is established.

Our approach does not require this assumption and is therefore better suited for modeling realistic cold atom systems, where steady states are difficult to maintain over a long time.
Nevertheless, since quasiparticles require a finite time to transition between reservoirs, we expect that our results should align with those obtained using the Landauer-B\"uttiker or Kubo formalisms over a certain intermediate time window, provided that the reservoirs are sufficiently large and a quasi-steady regime emerges. During this time window, the Landauer-B\"uttiker formalism is particularly valuable, as it provides frequency-dependent transmission coefficients that offer useful insights into thermal transport. Therefore, as a complementary perspective, we also employ the Landauer-B\"uttiker formalism to compute thermal conductance in the same setup, showcasing its relevance for understanding our results. 

\subsubsection{Landauer-B\"uttiker formula approach}
In our system, both the Kitaev model and the transverse-field Ising model can be reformulated in terms of Majorana fermions [see Eqs.~(\ref{MajH}) and \eqref{eq:HR}]. However, when expressed in their respective eigenstate representations, the tunneling Hamiltonian involves processes that do not conserve the number of quasiparticles. To address this, we extend the analysis to the Nambu representation, where both particle and hole degrees of freedom are explicitly included, and re-derive the Landauer-B\"uttiker formula to accurately capture the transport properties. The resulting energy current is given by (see App.~\ref{app: Landauer}):
\begin{equation}
    J_E=\frac{1}{2}\int_{-\infty}^{\infty}\frac{\mathrm d\omega}{2\pi}\omega\mathcal{T}(\omega)[n_F(\omega,T_l)-n_F(\omega,T_r)].\label{eq:Landauer}
\end{equation}
Here, $n_F(\omega,T)$ is the Fermi distribution at frequency $\omega$, temperature $T$, and zero chemical potential; and 
\begin{equation}
    \mathcal{T}(\omega)={\rm Tr}[G^R(\omega)\Gamma^r(\omega)G^A(\omega)\Gamma^l(\omega)] , \label{eq:trans}
\end{equation}
is the energy-dependent transmission rate. Here, $G^R(\omega)$ and $G^A(\omega)$ are the retarded and advanced Green functions in the Nambu representation, respectively, and $\Gamma^{l/r}(\omega)$ denotes the imaginary part of the self-energy associated with the left and right reservoirs. 
In the derivation, the key assumption is that the Green functions of the reservoirs do not change throughout the evolution. This assumption holds when the number of sites in the reservoirs is sufficiently large. As a result, we expect the Landauer-B\"uttiker formula to yield results that are consistent with the direct numerical simulations when the reservoirs are sufficiently long.

Given the thermal current $J_E$, the corresponding thermal conductance follows directly as:
\begin{equation}
    \kappa/T=2J_E/|T_l^2-T_r^2|, \label{eq:kappaLB}
\end{equation}
along the same lines as in Eq.~\eqref{eq:kappalr}.

\section{\label{sec:measure}Measuring the thermal conductance and analysis of contributing factors via Landauer-B\"uttiker formalism}

Having introduced the system configuration and computational methods, we now present our detailed results on the buildup of quasi-thermal conductance and then compare these with the results obtained from the Landauer-B\"uttiker formula. In our setup, quasi-thermal/thermal conductance is not measured directly but is inferred from the thermal current divided by the temperature difference between the reservoirs, normalized by their average temperature [see Eqs.~\eqref{eq:kappalr} and \eqref{eq:kappaLB}]. This indirect definition requires the temperature bias to be sufficiently small to ensure the validity and stability of the extracted conductance. In our calculations, we fix the specific ratio $T_l/T_r\!=\!1.2$, with the Kitaev channel initially set to $(T_l + T_r)/2$. This choice ensures that the temperature difference remains sufficiently small for the extracted conductance to be meaningful and stable (see App.~\ref{DeltaT} for a justification of this choice).

\subsection{Numerical study and validity of the Landauer-B\"uttiker formula}
\begin{figure}[t!]
    \centering
    \includegraphics[width=0.48\textwidth]{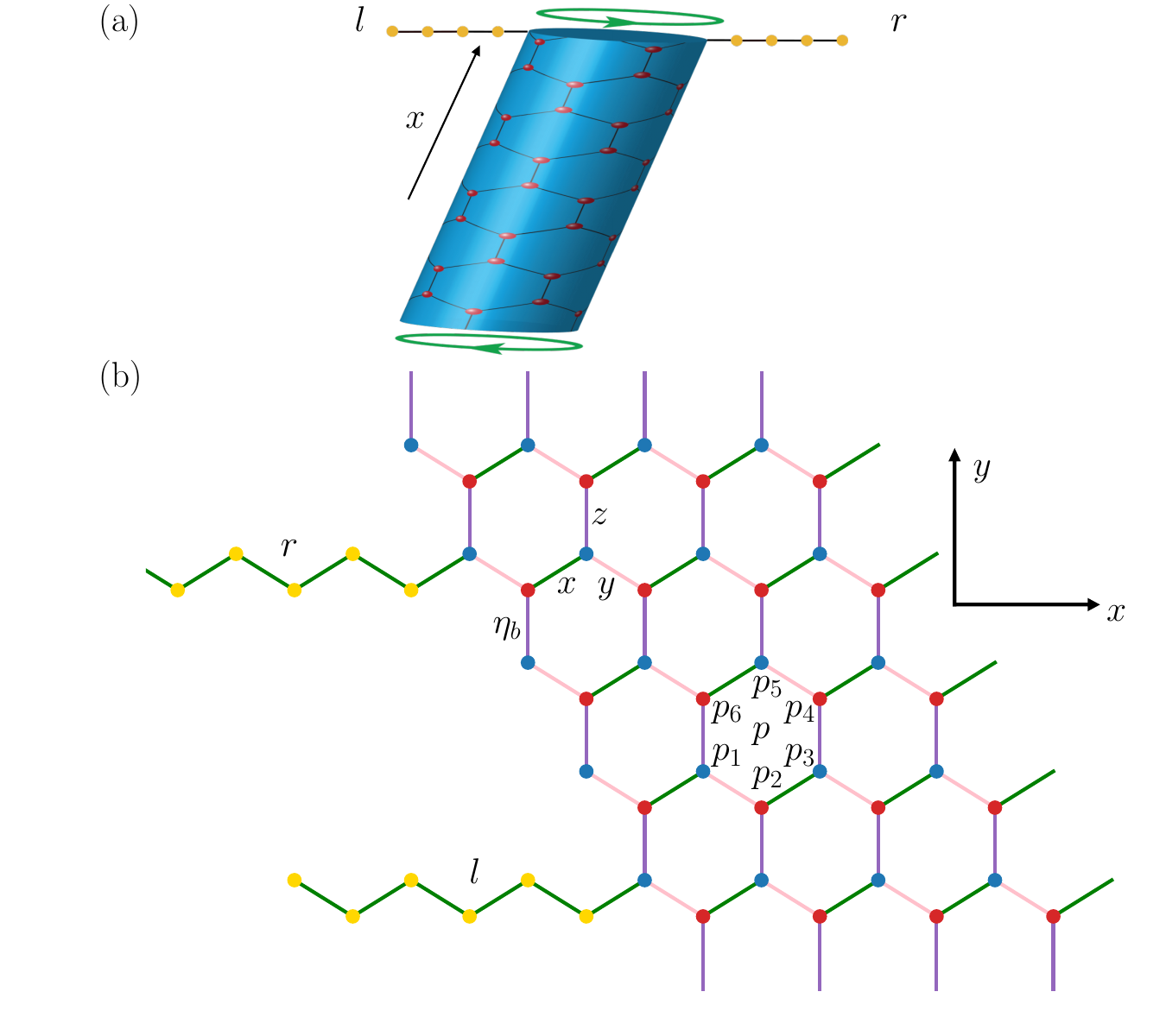}
    \caption{\textbf{Structure of the system used in numerical calculation.} (a) The overall configuration of the system, and (b) its enlarged view. The central channel, a cylindrical chiral spin liquid, has open boundaries in the $x$ direction and periodic boundaries along $y$. The two reservoirs (yellow sites), each formed by a one-dimensional transverse-field Ising model with open boundaries, connect to the channel at diametrically opposed locations along the same side of the $y$-direction (i.e., for a system with $L_y$ unit cells along the periodical direction, the reservoirs are connected to the $y=0$ and $y=L_y/2$ points). This specific figure (b) displays an example with six unit cells along $y$. The label $l/r$ in the figure denotes the left and right reservoirs, respectively. The green ($x$), pink ($y$), and purple ($z$) bonds are the same as those shown in Fig.~\ref{Fig1}.}
    \label{Fig1_2}
\end{figure}

We now numerically explore the scheme introduced in Sec.~\ref{sec:Model}. For calculation purposes, we impose periodic boundary conditions along the $y$-direction, resulting in a cylindrical geometry for the central Kitaev system, to which the reservoirs are coupled (see Fig.~\ref{Fig1_2})~\cite{lacki2016quantum}. In our calculations, we consider a system comprising sixteen sites along the periodic $y$-direction and eight unit cells along the open boundary $x$ direction. The reservoirs, connected at opposite points along the same edge (see Fig.~\ref{Fig1_2}), consist of 70 to 200 lattice sites depending on the case under study. With this configuration, we first demonstrate how a stable thermal conductance can be extracted from the direct time evolution simulation and then compare the results with those obtained from the Landauer-B\"uttiker formalism.

\begin{figure}[t!]
    \centering
    \includegraphics[width=0.49\textwidth]{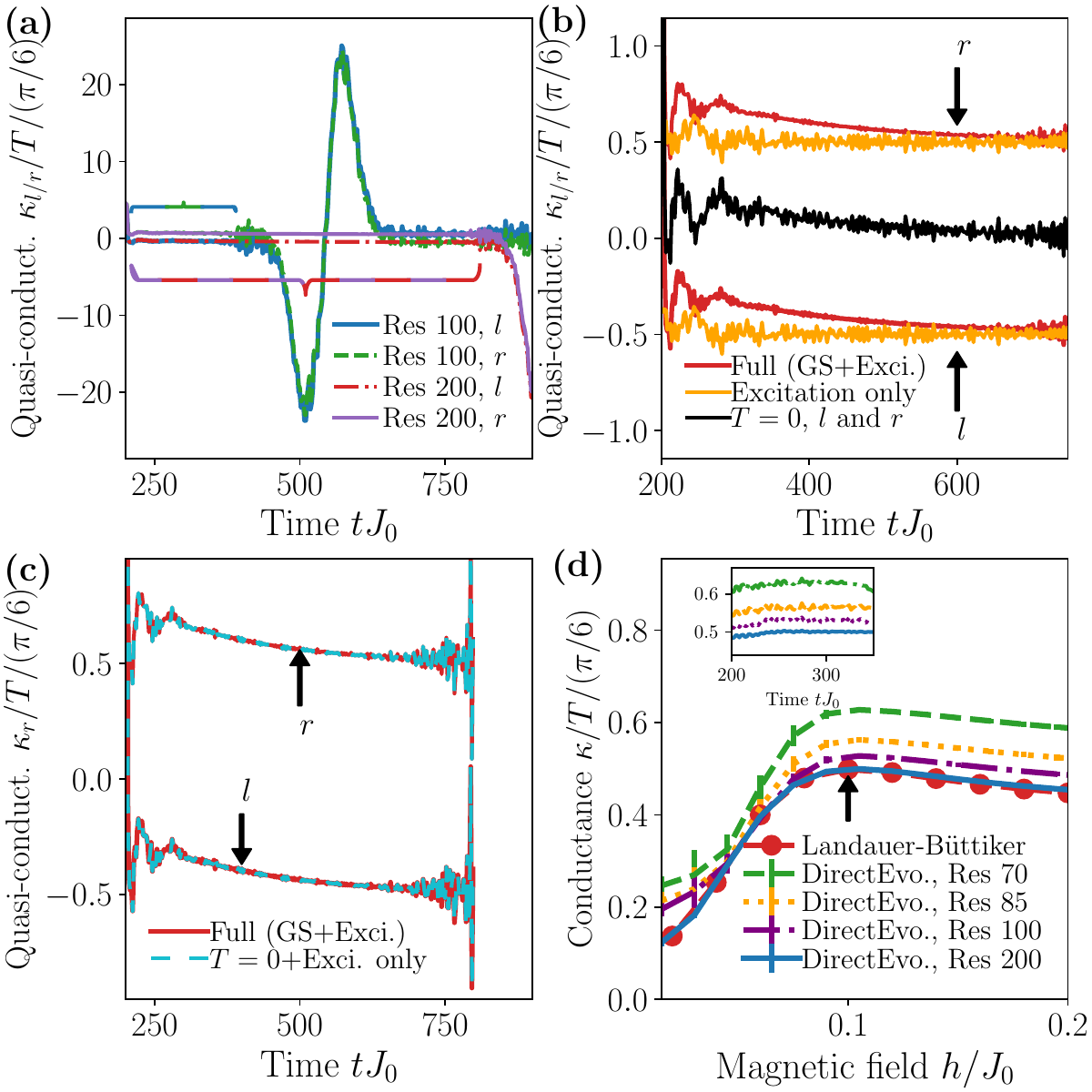}
    \caption{\textbf{Extracting thermal conductance from time-evolved quasi-thermal conductance and comparison with Landauer-B\"uttiker results.} \textbf{(a-c)} Time evolution of quasi-thermal conductance. Panel (a) shows systems with 100 and 200 reservoir sites, labeled as "Res 100" and "Res 200" for brevity. For both cases, quasi-stable regions appear as marked by the alternating blue-and-green curly bracket
    for "Res 100" (blue and green curves) and the alternating red-and-purple curly bracket
    for "Res 200" (red and purple curves), with the length of these quasi-stable regions dependent on the reservoir size.
    Panel (b) shows the quasi-thermal conductance for reservoirs of length 200 sites under three conditions: the red curves include both ground-state and excitation contributions [Full (GS+Exci.)], the black curve shows only the ground-state contribution at zero temperature ($T = 0$), and the orange curves include only the excitation contributions [Marked as "Excitation only", which means the last term of Eq.~\eqref{eq:rho_used} is removed]. 
    When the ground-state term is removed, the conductance becomes symmetric between the two reservoirs, reflecting the thermal conductance due solely to excitations--the primary focus of this work.
    Panel (c) demonstrates that the full quasi-thermal conductance (red) is exactly reproduced by the sum of the zero-temperature conductance and the excitation-only conductance (cyan).
    \textbf{(d)} Magnetic field dependence of thermal conductance from direct time evolution calculations (DirectEvo.) and the Landauer-B\"uttiker formula, showing strong agreement for long reservoirs (blue curve), with the arrow indicating the value of $h$ used in panels (a)-(c). The inset figure shows the quasi-thermal conductance for each reservoir length at $h=0.1J_0$. "$l$" and "$r$" denote the left and right reservoirs, respectively. All the results in this figure are normalized by the quantized thermal conductance $\pi/6$. 
    System size: $8_x \times 16_y$, 200 reservoir sites; parameters: $h = 0.1J_0$, $J_T = 0.8J_0$, $T_l = 0.006J_0$, $T_r = 0.005J_0$, unless otherwise specified. }    
    \label{Fig2}
\end{figure}

\subsubsection{Thermal conductance from direct time evolution simulation}

As introduced in the previous section, the time evolution of the quasi-thermal conductance $\kappa_{l/r}/T$ [see Eq.~\eqref{eq:kappalr}] can be obtained numerically and the result after the ramping, i.e., for $t>10t_s=200J_0$, is shown in Fig.~\ref{Fig2}(a). The curves in different colors correspond to different reservoir lengths, with longer reservoirs shown in purple and red, and shorter ones in green and blue. The plot reveals a smooth ``plateau" region in the conductance, observed from $t$  slightly above $200J_0$ to $800J_0$ 
in the purple and red curves, as shown by the alternating red-and-purple curly bracket
 and from $t$ slightly larger than $200J_0$ to $400J_0$ in the green and blue curves as shown by the alternating blue-and-green curly bracket.
 This region is expected to correspond to the quasi-steady transport window, during which excitations propagate between the two reservoirs in a relatively stable manner, i.e., after they have passed the initial transient dynamics but before they are significantly depleted in the reservoirs. It is within this interval that the real thermal conductance can be reliably extracted.

The duration of the smooth ``plateau" region depends on the reservoir lattice size. As shown in Fig.~\ref{Fig2}(a), when the reservoir lattice size decreases from 200 sites (purple and red) to 100 (green and blue), the smooth region duration shortens accordingly. This is because the number of available excitations in the reservoirs scales with the lattice size. With fewer sites, fewer excitations are supplied, and those in the shorter reservoir are exhausted earlier, leading to a shorter plateau.

However, it is important to note that this smooth region in the quasi-thermal conductance cannot be directly interpreted as the actual thermal conductance of the system. To illustrate this more clearly, we zoom into the smooth region and plot it as red curves in Fig.~\ref{Fig2}(b). It is evident that the quasi-thermal conductance on the left and right reservoirs is asymmetric with respect to zero. This indicates that no heat current is conserved between the two reservoirs, with the additional energy leaking from the channel into the reservoirs. In fact, this leaked heat originates from the influence of the ground state
[see the last term of Eq.~\eqref{eq:rho_used}]. 

To further illustrate the impact of the ground state,  we compute the time-dependent quasi-thermal conductance at zero temperature [i.e., setting $n_F(\xi_k,T)=0$ in Eq.~\eqref{eq:rho_used}], shown as the black curve in Fig.~\ref{Fig2}(b). In this case, the entire system is at zero temperature, and no excitations are present. Consequently, the observed conductance reflects only the ground-state contribution, corresponding to the last term in Eq.~\eqref{eq:rho_used}. Due to the symmetry between the two reservoirs, their conductance remains identical, exhibiting a slow decay. Upon subtracting the ground-state contribution, i.e., removing the last term in Eq.~\eqref{eq:rho_used},  the quasi-thermal conductance at $T_l = 0.006J_0$ [orange curve in Fig.~\ref{Fig2}(b)] exhibits plateau regions in the curves of the left and right reservoirs that are symmetric about zero~\footnote{The fluctuations of the orange signal lack perfect symmetry, which is expected, as it takes time for variations in one reservoir to propagate to the other.}. This behavior aligns with the transport properties predicted by the Landauer-B\"uttiker formalism~\cite{haug2008quantum}, where conductance arises from excitations.

This observation highlights that the asymmetry and slow decay in the full numerical results [red curve in Fig.~\ref{Fig2}(b)] stem from the ground-state contribution. Further confirmation is provided in Fig.~\ref{Fig2}(c), where the sum of the $T=0$ result and the excitation-only contribution (cyan curve) precisely reproduces the exact result (red curve). Moreover, at long times after the ramping process (for example, $tJ_0>600$), the quasi-thermal conductance including all contributions [red curves in Fig.~\ref{Fig2}(b)] converges to the result obtained when the ground-state contribution is omitted (orange curves). This is expected: in the limit of infinitely long reservoirs, where excitations are never depleted, a genuine steady state can be established in the long-time limit, as is typically assumed in transport studies. In this case, the ground-state contribution vanishes, so that only excitation-based transport governs the conductance. 

Since large reservoirs are typically difficult to implement in quantum-engineered platforms, it is desirable to identify a method that systematically suppresses the influence of the ground state. Here, we use the fact that the two reservoirs are identical and thus have equal ground-state contributions to the thermal current. Therefore, by taking the difference of the quasi-thermal conductance of the two reservoirs, these ground-state effects cancel out. Accordingly, the genuine thermal conductance associated with the excitations is calculated as half of this difference, i.e.,
\begin{equation}
\kappa/T=|(J^l_E-J^r_E)/(T_l^2-T_r^2)|. \label{eq:kappa}
\end{equation}
Under this setup, a more stable region appears, as shown in the inset of Fig.~\ref{Fig2}(d). We can fit these regions by a horizontal function $(\kappa/T)$ across different magnetic field strengths, with the obtained $\kappa/T$ illustrated in Fig.~\ref{Fig2}(d). 

We note that the time required to establish a stable plateau can be estimated from the group velocity of the edge states, yielding $4\pi L_y/(9 \sqrt{3}h)$, with $L_y$ being the number of unit cells along the $y$-direction~\cite{sun2023engineering}. For $h=0.05J_0$ as an instance, its value can be calculated to be $258/J_0$. In practice, we select the fitting window $tJ_0=400$ to $600$ for the case with 200 reservoir sites, ensuring that the plateau is fully developed across all cases considered. For a typical Rydberg-atom platform with $J_0=4$MHz~\cite{chen2024realization}, this corresponds to a real-time interval of roughly 100-150~$\mu$s, which is experimentally accessible.

Although error estimates due to small fluctuations are included [see the inset of Fig.~\ref{Fig2}(d)], they are nearly imperceptible in the figure due to their minimal size under the given parameters. This consistency further supports the validity of describing the system's thermal conductance as half the difference in conductance between the two reservoirs. 

Another property shown in Fig.~\ref{Fig2}(d) is that the obtained thermal conductance decreases with increasing the number of sites in the reservoir, eventually reaching a stable value once the reservoir site number is sufficiently large. Since the transport properties of the channel should be unaffected by the length of the reservoirs, we consider the results obtained with sufficiently large reservoirs to provide a reliable measure of the system's thermal conductance.

\subsubsection{Comparison with the Landauer-B\"uttiker formalism}

Beyond the direct time-evolution calculations based on Eq.~\eqref{eq:kappa}, the system can also be analyzed using the Landauer-B\"uttiker formula; see Refs.~\cite{wan2024quarter,yan2021a,huang2018disorder} for previous applications to Majorana-related transport problems. Using Eq.~\eqref{eq:kappaLB}, we evaluate the thermal conductance from the Landauer-B\"uttiker formula and compare it with the results obtained from the direct time-evolution calculation [Eq.~\eqref{eq:kappa}]. 
Importantly, the Landauer-B\"uttiker approach captures transport solely between quasiparticle excitations; hence, the results are independent of the ground state contribution. Remarkably, as shown in Fig.~\ref{Fig2}(d), the Landauer-B\"uttiker-derived conductance (red curve) aligns closely with the direct time-evolution calculation if the length of the reservoir is large enough (blue curve). This is understandable because a sufficiently long reservoir satisfies the assumption of the Landauer-B\"uttiker formalism, where the Green function in the reservoir remains nearly unchanged during the detection process. This agreement highlights the reliability of the Landauer-B\"uttiker formula in this context, and hence its utility in explaining the observed transport phenomena.

\subsection{\label{InfluenceJTandT} Influence of coupling strength and temperature on transmission rate}
Having established the validity of the Landauer-B\"uttiker formalism in capturing the transport properties of our system, we now employ it to analyze the underlying physical mechanisms and examine how the thermal conductance depends on various system parameters. Given the practical constraints in quantum-engineered platforms, where the total number of lattice sites is typically limited, we focus on systems comprising a few hundred sites. In this regime, finite-size effects become prominent, resulting in a discrete energy level structure in the central channel [e.g., yellow lines in Fig.~\ref{Fig4}(a) and (b) indicate the location of the discrete energy levels]. In addition, finite-size effects also give rise to a finite bulk gap even at zero magnetic field as shown by Fig.~\ref{fig:bulkgap} in App.~\ref{app:bulkgap}, where the bulk gap is shown to increase with the magnetic field $h$ and decrease with the number of sites along the $y$-direction. In our transmission results, any peak that appears within the bulk gap (e.g., between the two black dotted lines in Fig.~\ref{Fig4}, which mark the bulk gap edges) originates from contributions of the chiral edge states.

On the other hand, coupling to the reservoirs introduces a finite lifetime to these levels, manifesting as a broadening characterized by the imaginary part of the self-energy, denoted as $\Gamma^{l/r}$, as used in Eq.~\eqref{eq:trans}.
The interplay between this level broadening and the intrinsic level spacing gives rise to a rich variety of transport behaviours. To explore this, we systematically vary the reservoir-channel coupling strength $J_T$, thereby tuning the effective broadening $\Gamma^{l/r}$. This allows us to investigate how the relationship between $\Gamma^{l/r}$ and the typical energy level spacing influences the transmission spectrum. 

\subsubsection{\label{sec:InfCoup}Influence of coupling strength}
To systematically explore different transport regimes, we analyze the transmission spectrum $\mathcal{T}(\omega)$ by varying the coupling strength $J_T$. The interplay between the broadening $\Gamma^{l/r}\propto J^2_T$ and the discrete energy level structure of the finite-sized channel leads to qualitatively distinct behaviors, depending on whether $\Gamma^{l/r}$ is smaller than, comparable to, or larger than the typical spacing between nearby levels.

\begin{figure}[]
    \centering
    \includegraphics[width=0.49\textwidth]{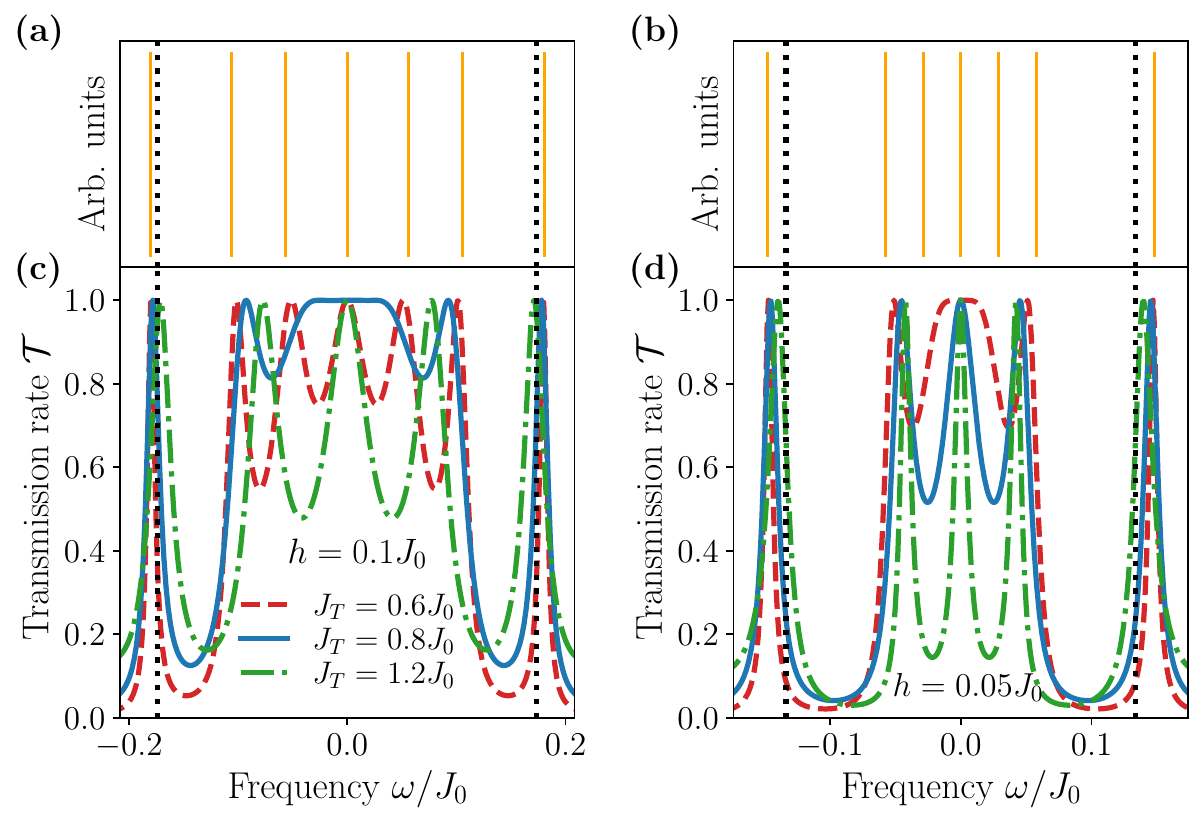}
    \caption{
\textbf{Discrete central channel energy levels and frequency dependence of the transmission rate under different magnetic fields and coupling strengths.}
\textbf{(a)} and \textbf{(b)} Yellow lines mark the discrete energy levels of the central channel in frequency space [$x$-axis, shared with panel (c)], for $h=0.1J_0$ and $0.05J_0$, respectively.
\textbf{(c)} At $h=0.1J_0$, a near-unity transmission plateau emerges at $J_T=0.8J_0$, corresponding to the optimal $J_T^\ast$ in Fig.~\ref{Fig3}(a). This plateau enables half-quantized thermal conductance, whereas deviations from $J_T^\ast$ remove it.
\textbf{(d)} At $h=0.05J_0$, with the same $J_T$ values as in (c), the optimal coupling shifts to $J_T^\ast=0.6J_0$, yielding a narrower plateau.
The Kitaev channel dimensions are $8_x \times 16_y$, and each reservoir has 200 sites. The black vertical dotted lines mark the bulk gap edges $\pm\Delta$ of the finite system, and transmission peaks appearing within $(-\Delta,\Delta)$ originate from edge states.
}
    \label{Fig4}
\end{figure}

Figure~\ref{Fig4}(c) shows $\mathcal{T}(\omega)$ for $h = 0.1J_0$ at various $J_T$. In the weak coupling regime ($J_T = 0.6J_0$, red curve), $\Gamma^{l/r}$ is appreciably smaller than the energy level spacing, and the transmission spectrum features well-separated sharp peaks that align with the discrete eigenvalues of the channel shown by the yellow lines in Fig.~\ref{Fig4}(a). This regime reflects a situation where states are weakly hybridized with the reservoirs, preserving their identity as isolated modes.

Notably, even in this regime, the valleys adjacent to the central peak at $\omega = 0$ maintain finite transmission values, which increase gradually with $J_T$ until the optimal coupling $J_T^\ast = 0.8J_0$ (blue curve), where the zero-energy mode and its neighboring levels broaden sufficiently to merge into a plateau of nearly unit transmission across a finite frequency window. This broad, high-transmission feature enhances the robustness of half-quantized thermal conductance at finite temperatures by ensuring that the dominant excitations fall within the plateau region, as discussed in the following subsection. The emergence of a perfect transmission plateau at optimal coupling is a well-established feature of two-terminal transport setups with a finite-size central region hosting chiral edge modes~\cite{gavensky2021nonequilibrium}.  We further illustrate in App.~\ref{app: few-sites} that its width can be estimated by the energy spacing between the two lowest eigenstates.

As the coupling strength is increased beyond the optimal value, the trend continues. At stronger coupling ($J_T = 1.2J_0$, green curve), $\Gamma^{l/r}$ becomes significantly larger than the level spacing. This detuning from the optimal coupling strength causes the plateau to collapse into a narrower peak centered at $\omega = 0$, reducing the effective energy range over which $\mathcal{T} \approx 1$, as illustrated by the toy model in App.~\ref{app: few-sites}. This non-monotonic evolution of the transmission profile with coupling strength is characteristic of finite-size systems and reflects the sensitivity of the transport to the relationship between $\Gamma^{l/r}$ and the level spacing.

The same qualitative behavior is observed at a lower magnetic field $h = 0.05J_0$, as shown in Fig.~\ref{Fig4}(d). Due to the reduced spectral gap, which depends linearly on the magnetic field strength $h$, the level spacing tends to decrease as well. This decrease consequently shifts the optimal coupling to a smaller value, $J_T^\ast = 0.6J_0$, and the resulting transmission plateau becomes narrower. Increasing $J_T$ beyond this optimal value again leads to the collapse of the plateau into a sharp peak, mirroring the behavior seen at higher magnetic fields. These results indicate that the value of $J_T^\ast$ required to realize a broad near-unity transmission plateau depends sensitively on the magnetic field and the underlying level structure of the channel.

Overall, this analysis highlights how the interplay between intrinsic level spacing and reservoir-induced broadening governs the transmission rate in mesoscopic topological systems. However, the measured thermal conductance is further influenced by the excitation distribution, which depends on the temperature. In the following, we discuss how temperature affects the thermal conductance. 

\subsubsection{Influence of temperature \label{sec:inf_temp}}
According to the Landauer-B\"uttiker formalism [Eq.~\eqref{eq:Landauer}], the thermal conductance is obtained by integrating $\mathcal{T}(\omega)$ against the difference in occupation functions. In particular, by setting $\mathcal{T}(\omega)=1$ in Eq.~\eqref{eq:Landauer}, one obtains the result $\kappa/T = \pi/12$  (in natural units with $k_B = \hbar = 1$), which corresponds to the half-quantized value~\cite{nasu2017thermal,sumiyoshi2013quantum}. Therefore, to achieve half-quantized thermal conductance, $\mathcal{T}(\omega)$ must remain close to unity over the frequency range where the excitation distribution, i.e., the Fermi distribution $n_F(\omega,T)$ in Eq.~\eqref{eq:Landauer}, is appreciable, which is determined by the temperature. 

This condition is naturally satisfied at low temperatures, where the excitation distribution becomes sharply localized near $\omega = 0$. In this regime, even if the transmission plateau is narrow or imperfect due to non-optimal tunnel coupling $J_T$ or magnetic field $h$, the dominant excitations remain confined within the high-transmission region. As a result, the system can still robustly exhibit half-quantized thermal conductance. Lowering the temperature effectively reduces the sensitivity to the edges of the plateau, thereby enhancing the stability of the half-quantized signal. 

This robustness is rooted in the presence of a topologically protected zero-energy Majorana mode in our system configuration, which guarantees $\mathcal{T}(\omega)\to 1$ as $\omega\to 0$ (see App.~\ref{app:LB_half}). This mode originates from the nontrivial topology and the particle-hole symmetry of the chiral spin liquid in the Majorana representation~\cite{kitaev2006anyons}. Note that we have verified the topological nature of our finite-size sample by calculating the many-body Chern number via twisted boundary conditions (see App.~\ref{app:Chern})~\cite{niu1985quantized,nathan2024relating}, yielding the Chern number $C = \pm 1$ depending on the sign of the magnetic field $h$.

At higher temperatures, however, the excitation distribution broadens over a larger frequency window, and the system becomes more sensitive to deviations of $\mathcal{T}(\omega)$ from unity beyond the center of the plateau. As a result, even small imperfections or narrowing in the transmission spectrum can reduce the final conductance $\kappa/T$ from its half-quantized value.

Finally, it is important to emphasize that the pronounced temperature dependence arises from finite-size effects. These effects cause transmission valleys to appear between eigenstates, except at the optimal coupling $J_T^\ast$ (see Fig.~\ref{Fig4}). As a result, although multiple edge states exist in the system, only the lowest-energy state, i.e., the zero-energy state, can be effectively targeted to realize half-quantized thermal conductance, except at $J_T^\ast$, where the second-lowest eigenstate also contributes. On the other hand, as discussed in App.~\ref{LargerSystem}, increasing the system size produces a denser energy spectrum and a smoother transmission function $\mathcal{T}(\omega)$. This relaxes the low-temperature requirement and allows near half-quantized conductance to persist at higher temperatures (of the order of the topological bulk gap). These results connect finite-size physics with the conventional picture of chiral edge transport in large topological systems.

\section{\label{sec:num_thermal}Achieving half-quantized thermal conductance: parameter dependence}
Having examined how parameters, such as tunnel coupling and magnetic field, shape the transmission function and discussed how temperature influences the reservoir excitation spectrum, we now explore how these factors together determine the emergence and stability of half-quantized thermal conductance in the entire system. Specifically, we examine the dependence of the thermal conductance on these parameters and identify the regimes in which the half-quantized value emerges.
\subsubsection{Dependence on coupling strength, magnetic field, and temperature}
\begin{figure}[]
    \centering
    \includegraphics[width=0.49\textwidth]{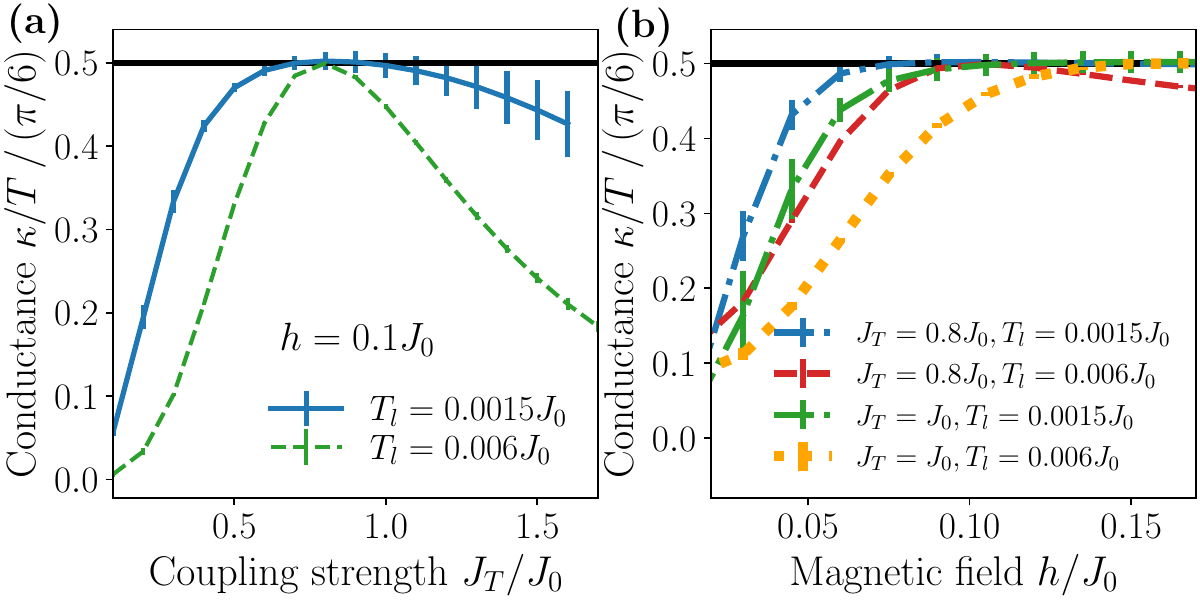}
    \caption{\textbf{Coupling strength and magnetic field dependence of thermal conductance.} 
    \textbf{(a)} Thermal conductance as a function of coupling strength $J_T$ obtained from the direct numerical simulation [see Eq.~\eqref{eq:kappa}] at different temperatures (see legend). The results are normalized by the quantized thermal conductance $\pi/6$ and the black horizontal line marks the half-quantized thermal conductance.
    There is an optimal coupling strength $J^\ast_T \approx 0.8 J_0$ at $h=0.1J_0$, where conductance reaches the half-quantized value, with lower temperatures expanding this region. 
    \textbf{(b)} Thermal conductance as a function of the magnetic field across different temperatures and coupling strengths, showing a plateau at the half-quantized value under very low temperatures. The system size is  $8_x\times16_y$ for the chiral spin liquid channel, $200$ reservoir sites at $T_l=0.006J_0$, and $600$ sites at $T_l=0.0015J_0$. }
    \label{Fig3}
\end{figure}

Building on the understanding established via the Landauer-B\"uttiker analysis, we now examine the thermal conductance calculated from the direct time-evolution calculation (see Sec.~\ref{sec:exact}). As shown in Fig.~\ref{Fig3}(a) with $h=0.1J_0$, which is the same field used in Fig.~\ref{Fig4}(c), the conductance depends sensitively on the tunnel coupling $J_T$ between the chiral spin liquid channel and the reservoirs. At the optimal coupling $J_T^* \approx 0.8 J_0$, where the transmission rate exhibits a plateau as shown by the blue curve in Fig.~\ref{Fig4}(c), the thermal conductance approaches the half-quantized value $\pi/12$ (black line)~\cite{nasu2017thermal,sumiyoshi2013quantum}.

While deviating from the optimal value tends to reduce the conductance below the half-quantized value, lowering the temperature helps broaden the parameter regime in which the conductance remains close to this value. For instance, as the reservoir temperature decreases from $T_l = 0.006J_0$ (green) to $T_l = 0.0015J_0$ (blue), the half-quantized region becomes more pronounced, though oscillations in the signal may increase, introducing greater fitting uncertainty. This trend is consistent with the fact that lower temperatures reduce sensitivity to the edges of the transmission plateau, as discussed in Sec.~\ref{sec:inf_temp}. These observations demonstrate that even in finite systems, robust half-quantized conductance can be achieved over a finite window of $J_T$, and this window expands as temperature approaches zero. Importantly, we should note that at lower temperatures, the number of reservoir sites must be increased (e.g., from 200 to 600) to ensure sufficient excitation density; otherwise, the reservoirs fail to approximate a thermal steady state, and the Landauer-B\"uttiker picture no longer applies [as illustrated in Fig.~\ref{Fig2}(d)].

In Fig.~\ref{Fig3}(b), we further explore the thermal conductance as a function of magnetic field $h$ for different temperatures and coupling strengths. At $T_l=0.006J_0$ and $J_T=0.8J_0$ (red curve), the conductance initially increases with $h$ but subsequently decreases without forming a plateau. In contrast, at sufficiently low temperatures such as $T_l=0.0015J_0$ (blue curve), a plateau emerges at the half-quantized value. Similarly, for $J_T = J_0$, lowering the temperature from $T_l = 0.006J_0$ (yellow curve) to $T_l = 0.0015J_0$ (green curve) expands the half-quantized region. It is noted that the plateau appears with the field-induced spectral gap significantly exceeding the thermal energy. For example, at $T_l = 0.0015J_0$, a plateau emerges when $h > 0.07J_0$, corresponding to a topological bulk gap larger than $0.09J_0$ (in the thermodynamic limit), which is nearly 60 times the reservoir temperature. These observations indicate that ultracold conditions are required to observe quantized conductance in finite systems, where the transmission rate $\mathcal{T}(\omega)$ may fall below unity away from $\omega = 0$. As discussed earlier, lowering the temperature helps mitigate this issue by reducing the energy window within which excitations are thermally populated.

\subsubsection{Enhancing quantized conductance at higher temperatures via finite-size effects}
While our earlier analysis emphasized the necessity of ultralow temperatures for observing half-quantized thermal conductance, primarily due to finite-size-induced suppression of transmission away from $\omega = 0$, a notable exception emerges under special conditions. Specifically, we find that the half-quantized thermal conductance can emerge at higher temperatures when the coupling strength is appropriately tuned.

\begin{figure}[]
    \centering
    \includegraphics[width=0.47\textwidth]{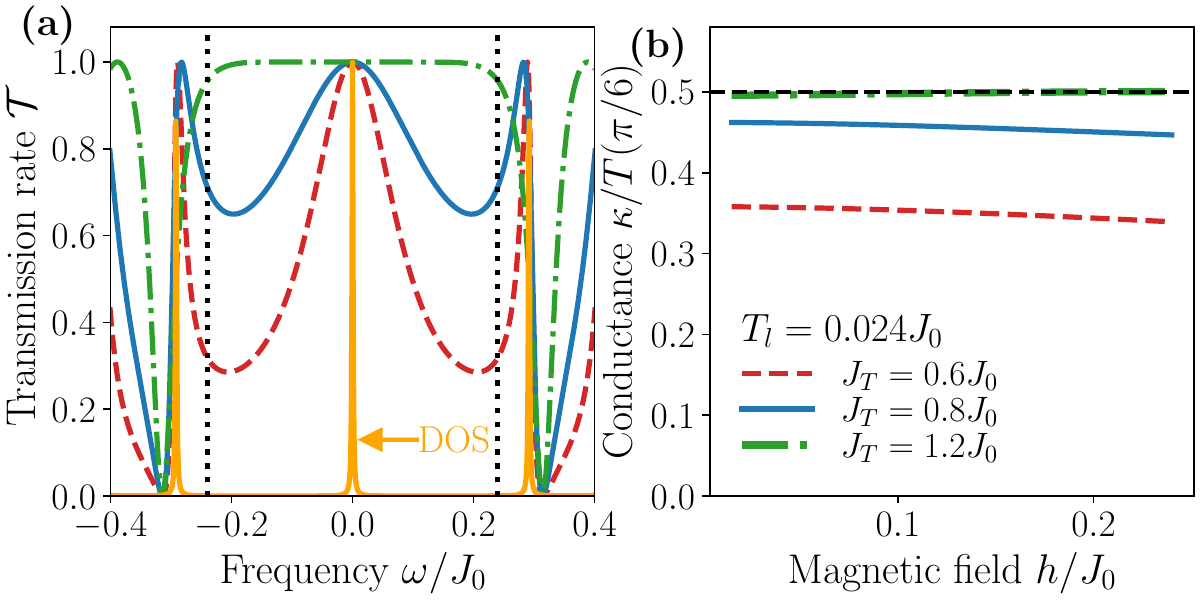}
    \caption{\textbf{Transmission rate and thermal conductance with four unit cells along the periodic direction}. \textbf{(a)} Transmission rate vs.~frequency under different coupling strengths at $h=0.1J_0$. With a limited number of sites, the large energy spacing between the lowest states results in a broad transmission plateau at optimal  $J^\ast_T$. The orange solid curves denote the density of states (DOS) of the central channel, normalized to a maximum value of one, with peaks indicating the positions of the eigenstates.  \textbf{(b)} Thermal conductance vs. magnetic field strength. Color and line styles match those in (a) for each coupling strength. A half-quantized thermal conductance plateau is observed up to $T_l=0.024J_0$ at the optimal coupling strength $J^\ast_T=1.2J_0$. The reservoirs have 200 sites in the calculation, and the dimension of the Kitaev channel is $8_x\times4_y$. The black vertical dotted line marks the location of the bulk energy gap $\Delta$ of the finite-size system. 
    }
    \label{Fig5}
\end{figure}

As illustrated in Fig.\ref{Fig5}(a), when the Kitaev channel contains only four unit cells along the periodic $y$ direction, the level spacing between the lowest energy eigenstates becomes significantly larger, as shown by the normalized density of states of the central channel (orange solid curves).  This increased spacing permits the formation of a broad transmission plateau near $\omega = 0$ under the optimal coupling strength since the width of the plateau is determined by the energy spacing between the lowest two states. For coupling strengths that deviate from the optimal value $J_T^\ast = 1.2 J_0$, as shown by the red and blue curves, the transmission plateau is suppressed. In contrast, at the optimal coupling (green curve), a flat and extended transmission plateau is centered around zero frequency. This behavior is supported by the simplified few-site model presented in App.~\ref{app: few-sites}.

This enhanced plateau directly enables robust half-quantized thermal conductance even at relatively high temperatures. As shown in Fig.~\ref{Fig5}(b), at $T_l = 0.024 J_0$, which is over an order of magnitude larger than the temperature scale used in Fig.~\ref{Fig3}, an almost flat thermal conductance plateau emerges at the half-quantized value at the optimal coupling. Importantly, even away from this optimal coupling, such as at $J_T = 0.8 J_0$, the conductance remains close to $90\%$ of the half-quantized value, confirming that this mechanism is not overly fine-tuned. These findings suggest that in cold-atom experiments, where system sizes are often limited, the half-quantized thermal conductance can be observed with more realistic, less extreme cooling conditions. We note that at such high temperatures, vortex excitations may begin to proliferate, as indicated by the quantum Monte Carlo results of Nasu et al.~\cite{nasu2017thermal}. The study of these vortex-related effects goes beyond the scope of this work.

\section{\label{sec:sum}summary}
We proposed a framework for observing half-quantized thermal conductance in finite-size quantum-engineered chiral spin liquids. By coupling a one-dimensional transverse-field Ising reservoir to a chiral spin liquid channel, we demonstrated that thermal transport can be measured via spin correlations at the reservoir-channel junction. Our direct numerical simulations reveal a stable thermal current region, even for finite reservoirs. By fitting half the thermal current difference between the reservoirs, we isolate the quasiparticle contributions while canceling ground-state effects. This approach yields a clear half-quantized plateau in the magnetic field dependence of the thermal current under low temperatures and optimal coupling parameters.

Our findings not only explore topological thermal transport in finite-sized systems but also provide a practical pathway for experimentally probing the exotic thermal behavior of Majorana quasiparticles in clean, tunable cold-atom setups. The half-quantized thermal conductance originates from topologically protected Majorana zero modes in the chiral spin liquid, which enforce that the transmission rate in Eq.~\eqref{eq:Landauer} approaches unity in the low-frequency limit. When reservoir excitations fall within this window, half-quantized thermal conductance emerges. In finite systems, however, the channel spectrum is discrete even within the bulk gap, resulting in transmission rates that appear as peaks located at the system's discrete energy levels rather than a flat profile. As a result, half-quantized conductance can only be observed at ultralow temperatures, well below the field-induced gap, ensuring that thermal excitations lie within the zero-frequency region of the transmission peak, where the transmission rate remains near unity. This in turn requires reservoirs long enough to provide energy levels in the relevant low-temperature window, so that sufficient excitations are available to sustain a quasi-steady transport signal. Nevertheless, coupling between the channel and reservoirs broadens these transmission peaks, and at an optimal coupling strength, the broadening maximizes the frequency range around zero with a unity transmission rate, allowing the half-quantized conductance to persist at higher temperatures. Therefore, a practical strategy is to first tune the coupling strength to find its optimal value and then lower the temperature under this optimal coupling, ensuring that the reservoirs are long enough, i.e., the results no longer depend on reservoir length, to identify half-quantized values in the magnetic-field dependence of the thermal conductance. These insights provide a roadmap for experimental implementation, outlining how to tune system parameters and what conductance signatures to expect, paving the way for future explorations in chiral-spin-liquid quantum simulators.

\textit{Acknowledgments:} B.Y.S. and Z.W.Z. acknowledge support from the National Natural Science
Foundation of China (Grants No. 12204399, No. 12474366, and No. 11974334), and Innovation Program for Quantum Science and Technology (Grant No. 2021ZD0301200). N.G. and B.B. acknowledge the financial support from the ERC (LATIS project), the EOS project CHEQS, the FRS-FNRS Belgium, and the Fondation ULB. L.P.G. acknowledges support provided by the FRS-FNRS Belgium and the L'Or\'eal-UNESCO for Women in Science Programme. M.B.~was funded by the Deutsche Forschungsgemeinschaft (DFG, German Research
Foundation) -- 563114538, and the European
Union (ERC, QuSimCtrl, 101113633). Views and opinions expressed are however those of the authors only and
do not necessarily reflect those of the European Union
or the European Research Council Executive Agency.
Neither the European Union nor the granting authority
can be held responsible for them.

\appendix

\section{\label{app:JW_spectrum} Jordan-Wigner transformation and Hamiltonian in Majorana representation}

Typically, the Jordan-Wigner transformation is expressed as
\begin{eqnarray}
\sigma^x_j&=&\prod_{j^\prime=1}^{j-1}(1-2n_{j^\prime})(a^\dagger_j+a_j),\\
\sigma^y_j&=&-i \prod_{j^\prime=1}^{j-1}(1-2n_{j^\prime})(a^\dagger_j-a_j),\\
\sigma^z_j&=&2n_j-1,
\end{eqnarray}
is employed for one-dimensional spin systems with open boundaries. Here $\sigma^\alpha_j=2S_j^\alpha$ are the Pauli matrices, and $a_j$ is the annihilation operator of the corresponding fermions. This transformation can be used to solve the spectrum of transverse-field Ising model reservoirs by further converting normal fermions into Majorana fermions via
\begin{eqnarray}
    c^{l/r}_j&=&a^{l/r}_j+a^{l/r\dagger}_j,  \\
    \bar{c}^{l/r}_j&=&i(a_j^{l/r\dagger}-a^{l/r}_j),
\end{eqnarray}
where the labels $l$ and $r$ indicate the spin at left and right reservoirs, respectively. 
Then, $H_R^{l/r}$ becomes
\begin{equation}
    H_R^{l/r}=\sum_j \frac{J_R}{4} i\bar{c}^{l/r}_jc^{l/r}_{j+1}-\frac{i B}{2} c^{l/r}_j\bar{c}^{l/r}_j.\label{eq:HR}
\end{equation}
The quasi-particle excitation spectrum of this system under the periodic boundary in the momentum space can be obtained as~\cite{nasu2017thermal}
\begin{equation}
    E(q)=2\sqrt{\frac{J_R^2}{16}+\frac{\cos q J_R  B}{4}+\frac{B^2}{4}}.
\end{equation}
Since the half-quantized thermal conductance mainly comes from the quasi-particles around $E_k=0$, $B$ must be half of $J_R$ to make this spectrum gapless.

Regarding our central Kitaev spin channel, given that the half-quantized thermal conductance primarily arises from states near zero energy, and for computational convenience, we adopt a cylindrical geometry in our calculations, as shown in Fig.~\ref{Fig1_2}. In this geometry, the lattice can be mapped onto a brick-wall structure, where the Jordan-Wigner transformation remains applicable~\cite{feng2007topological,chen2007exact,chen2008exact}. By adopting this approach, the exact solvable Hamiltonian of the chiral spin liquid can be obtained as~\cite{sun2023engineering}:
\begin{eqnarray}
        H_C&=&-\frac{i}{4}\sum_{j\in {\rm filled\; circles}} J(c_jc_{j_x}+c_jc_{j_y}+\eta_b c_jc_{j_z})\nonumber\\
        &&+\frac{ih}{8}\sum_p(c_{p_1}c_{p_3}+\eta_{b_2}c_{p_3}c_{p_5}+\eta_{b_1}c_{p_5}c_{p_1}\nonumber\\
        &&\qquad\quad +c_{p_4}c_{p_6}+\eta_{b_1}c_{p_6}c_{p_2}+\eta_{b_2}c_{p_2}c_{p_4}). 
        \label{MajH}
\end{eqnarray}
Here, the definitions are as follows: for sites $j$ belonging to one sublattice (denoted by red circles in Fig.~\ref{Fig1_2}), $c_j=i(-a_j+a^\dagger_j)$ and $\bar{c}_j=a_j+a^\dagger_j$; for sites belonging to the other sublattice (denoted by blue circles in Fig.~\ref{Fig1_2}), $c_j=a_j+a^\dagger_j$ and $\bar{c}_j=i(-a_j+a^\dagger_j)$. In this context, $p$ represents a single plaquette of the honeycomb lattice. The operator $\eta_b=i\bar{c}_j\bar{c}_{j_z}$ takes the value $\pm1$ on each $z$ bond $b=\langle jj_z\rangle_z$, with $b_1$ and $b_2$ calculated by $\langle p_1 p_6\rangle_z$ and $\langle p_3 p_4\rangle_z$ for each plaquette, respectively. It is noted that these $\eta_b$ determine the vortex of the system. In our calculation, we assume that the system is in the vortex-free sector, i.e., all $\eta_b=1$, since the temperature is very low.

Using these transformations, both the $H_T$ term and the thermal current can be calculated under the Majorana representation as
\begin{equation}
    H_T=\frac{J_T}{4} i\bar{c}_lc_{C_l}+\frac{J_T}{4} i\bar{c}_rc_{C_r}. 
\end{equation}

With all three Hamiltonian terms written in the Majorana fermion representation, we can formally write the total Hamiltonian as follows:
\begin{equation}
    H\equiv\frac{i}{4}\sum_{jj^\prime}A_{jj^\prime}(\eta) c_j c_{j^\prime}. \label{eq:A}
\end{equation}

\section{\label{app:thermal_current} Calculation of the thermal current}

With the Jordan-Wigner transformation, we can transform the spin form of the energy evolution Eq.~\eqref{eq:JE}, i.e., the energy current of the $l/r$ reservoir to the Majorana form as 
\begin{eqnarray}   
    &&\langle \dot{H}^{l/r}_R\rangle = i\langle[H(t),H^{l/r}_R]\rangle=-J_T  B \langle S^y_{l/r} S_{C_{l/r}}^x \rangle \nonumber\\
    =&&i 2 J_T  B {\rm Tr}[e^{i H(t) t}(c_{C_{l/r}} c_{l/r}-c_{l/r}c_{C_{l/r}})e^{-i H(t) t} \rho_0]\nonumber\\
    \equiv&&i 2 J_T  B{\rm Tr} [Q^\dagger(t) \rho Q(t)]. \label{eq:time_evo}
\end{eqnarray}
Here, $Q(t)=\mathrm e^{-tA}$ with $A$ being the matrix $A$ defined in Eq.~\eqref{eq:A}.  The elements of the density matrix $\rho$, denoted as $\rho_{ij}$ are obtained from $\rho_{ij}=\langle c^\prime_i c^\prime_j \rho_0\rangle$, with $\rho_0$ representing the initial density matrix of the system and $c^\prime$ denoting the Majorana Fermions at the channel and the reservoirs. To calculate this $\rho$ matrix, we need to represent it in the quasi-particle bases. Specifically, by diagonalizing the Hamiltonian, we can write the Hamiltonian under the eigenvectors' representations, which can be written as 
\begin{eqnarray}
    H^{l/r}_R&=&\sum_{k} E^\prime_{k} \beta^{l/r\dagger}_{k}\beta^{l/r}_{k},\\
    H_c&=&\sum_{k} \epsilon^\prime_{k} \gamma^\dagger_{k}\gamma_{k},
\end{eqnarray}
with the relationship $c^\prime_i=\sum_k\sqrt{2}U_{i,k}\alpha_{k}$. Here, $\alpha_{k}$ denotes $\beta^{l/r}_{k}$ and $\gamma_{k}$, $\sqrt{2}$ normalizes the operator $c$ since $\{c_i,c_j\}=2\delta_{ij}$, and $U^\dagger U$ being the identity matrix. For a system with $M$ unit cells and each unit cell has $2N$ sites, we set $k=-MN, -MN+1,...,-1$ for the holes and $k=1,2,..., MN$ for the particles. The obtained eigenvectors $\beta_k$ are actually written in the Nambu representation. Moreover, due to the particle-hole symmetry, we have 
$\beta^{l/r}_{k}=\beta^{l/r\dagger}_{-k}$, $\gamma_{k}=\gamma^\dagger_{-k}$, $E^\prime_{k}=-E^\prime_{-k}$, and $\epsilon^\prime_{k}=-\epsilon^\prime_{-k}$. Then, under the quasi-particle excitation bases, it becomes~\cite{nasu2017thermal}
\begin{eqnarray}
    H^{l/r}_R&=&\sum_{k>0} E_{k} \beta^{l/r\dagger}_{k}\beta^{l/r}_{k}-\frac{1}{2}\sum_{k>0} E_{k},\label{app:Hr}\\
    H_c&=&\sum_{k>0} \epsilon_{k} \gamma^\dagger_{k}\gamma_{k}-\frac{1}{2}\sum_{k>0} \epsilon_{k},\label{app:Hc}
\end{eqnarray}
with $\epsilon_{k}=2\epsilon^\prime_{k}$, $E_k=2E^\prime_k$, and the last terms of the two equations being the ground state energies of the channels and reservoirs, respectively.

With these definitions, the density matrix for the reservoirs can be written as 
\begin{eqnarray}
    \rho_{i,j}=\langle c^\prime_i c^\prime_j \rho_0\rangle=2\sum_k \langle  U^\dagger_{k,i} U_{j,k} \alpha^\dagger_{k}\alpha_{k} \rho_0\rangle.
\end{eqnarray}
Under the initial equilibrium state, we have 
\begin{equation}
\langle \alpha^\dagger_{k}\alpha_{k} \rho_0\rangle=\left\{\begin{array}{cc}
     n_F(\xi_{k},T) & k>0, \quad {\rm particles } \\
     1-n_F(\xi_{-k},T) & k<0,\quad {\rm holes }
\end{array}\right. \label{eq:Fermi_Exc}
\end{equation}
with $n_F(\xi_{k},T)=1/(e^{\xi_{k}/T}+1)$ being the Fermi distribution, $T$ being the temperature, and $\xi_{k}$ representing the corresponding $\epsilon_k$ or $E_k$. 
Further utilizing the relation $U_{ik}=U^\dagger_{-ki}$, we have 
\begin{equation}
    \rho_{i,j}=2\sum_{k>0}[(U^\dagger_{k i}U_{jk}-U_{ik} U^\dagger_{k j})n_F(\xi_k,T)+U_{ik}U^\dagger_{kj}].
\end{equation}
It is noted that the last term, $2\sum_{k>0}U_{ik}U^\dagger_{kj}$, is associated with the ground state energy, as it follows that $\langle H \rho_0 \rangle$ is just the ground state energy at zero temperature, i.e., the last terms in Eqs.~(\ref{app:Hr}) and (\ref{app:Hc}). When the reservoirs are connected to the channel, these terms contribute to the thermal current, independent of quasi-particle excitations' thermal transport. To eliminate their effect, we calculate half of the difference in quasi-thermal conductance between the two reservoirs, assuming the reservoirs are identical, as their contributions will then cancel symmetrically. Moreover, due to the calculation of the difference in the thermal current between the reservoirs, the temperature of the Kitaev channel does not influence the results; this is consistent with the Landauer-B\"uttiker formula where this temperature is also irrelevant.

\section{\label{app: Landauer} Details of the Landauer-B\"uttiker formula}
In this section, we show the detail of the Landauer-B\"uttiker formula used in this work, which is obtained following reference ~\cite{haug2008quantum}. Here, for simplicity, the reservoir is assumed to be sufficiently long such that its state remains approximately unchanged during the evolution. As shown in Eqs.~(\ref{app:Hr}) and (\ref{app:Hc}), we can write the Hamiltonian in the Nambu bases of $\beta_k$ and $\gamma_i$. Under this bases, the coupling term $H_T$ can be written as 
\begin{equation}
    H_T=\frac{J_T}{4}\sum_{\alpha=l,r} i\bar{c}_{\alpha}c_{C_{\alpha}}\equiv\sum_{kj\alpha}V_{kj}\beta^{\alpha\dagger}_{k}\gamma^\alpha_{j}+h.c. \label{eq:Vkpm}
\end{equation}
Here, $\beta$ and $\gamma$ have both the particle part and the hole part, $V_{kj}$ is obtained from the transformation of the Hamiltonian from the Majorana to the excitation bases, which fulfills
\begin{equation}
    V^\alpha_{k,j}=-V^{\alpha\ast}_{-k,-j}.
    \label{eq:V_Vconj}
\end{equation}
Now, we consider the energy transport of the quasiparticle excitations, the energy current at the left reservoir is defined as 
\begin{equation}
J_{l}=-\langle \dot{H}^{l}_R \rangle=-i\langle[H,\sum_{k} E^\prime_{k}\beta^{l\dagger}_{k}\beta^l_{k}\rangle.
\end{equation}
This can be obtained as 
\begin{equation}
    J_l=-i\langle \sum_{kj}-\bar{V}^l_{kj}E^\prime_{k}\beta^{l\dagger}_{k} \gamma_j+\bar{V}^{l\ast}_{kj}E^\prime_{k} \gamma^\dagger_j \beta^l_{k}\rangle.
\end{equation}
Here, $\bar{V}^\alpha_{kj}=2{V}^\alpha_{kj}$, with the factor 2 comes from $\{\beta^\alpha_{k},\beta^\alpha_{k^\prime}\}=\delta_{k,-k^\prime}$.

Define 
\begin{eqnarray}
    G^{l<}_{j,k}(t-t^\prime)=i\langle \beta^{l\dagger}_{k}(t^\prime)\gamma_j(t)\rangle,\\
    G^{l<}_{k,j}(t-t^\prime)=i\langle \gamma^\dagger_j(t^\prime)\beta^l_{k}(t)\rangle,
\end{eqnarray}
the thermal current as 
\begin{equation}
    J_l=\sum_{kj}\bar{V}^l_{kj}E^\prime_{k}G^{l<}_{j,k}(t,t)-\bar{V}^{l\ast}_{kj}E^\prime_{k}G^{l<}_{k,j}(t,t).
\end{equation}
For the time-ordered green function, we have 
\begin{eqnarray}
    &&-i\partial_{t^\prime}G^{lt}_{j,k}(t,t^\prime)=-i\langle \gamma_j(t)[H,\beta^{l\dagger}_{k}(t^\prime)]\rangle \nonumber \\
    &&= 2E^\prime_{k}G^{lt}_{j,k}(t,t^\prime)+\sum_{j^\prime}\bar{V}^{l\ast}_{kj^\prime}G^t_{j,j^\prime}(t,t^\prime).
\end{eqnarray}
Then, we have the contour-ordered Green functions
\begin{eqnarray}
    G^{lt}_{j,k}(t,t^\prime)=\sum_{j^\prime}\int dt_1 G^t_{j,j^\prime}(t,t_1)\bar{V}^{l\ast}_{kj^\prime}g^{lt}_{k}(t_1,t^\prime),
\end{eqnarray}
with $g^{lt}_{k}(t_1,t^\prime)$ being the bare green function in the reservoir, and
\begin{equation}
(-i\partial_{t^\prime}-E_{k})g^{\alpha t}_{k}(t_1,t^\prime)=\delta(t_1-t^\prime).
\end{equation}
Since the reservoir is sufficiently long and its internal states remain nearly unchanged, we can further assume that the system reaches a steady state. This allows us to express the equation in frequency space, and by using the properties of the contour-ordered Green function, we obtain:
\begin{equation}
    G^{l<}_{j,k}(\omega)=\sum_{j^\prime}\bar{V}^{l\ast}_{k,j^\prime}[G^R_{j,j^\prime}(\omega)g^{l<}_{k}(\omega)+G^<_{j,j^\prime}(\omega)g^{lA}_{k}(\omega)].
\end{equation}
Here, $G^R$ and $G^A$ are the retarded and advanced Green functions, respectively.
Similarly,
\begin{equation}
    G^{l<}_{k,j}(\omega)=\sum_{j^\prime}\bar{V}^l_{kj^\prime}[g^{lR}_{k}(\omega)G^<_{j^\prime j}(\omega)+g^{l<}_{k}(\omega)G^A_{j^\prime j}(\omega)].
\end{equation}
Further using the Keldysh equation 
\begin{equation}
    G^{\stackrel{>}{<}}(\omega)=G^R(\omega)\Sigma^{\stackrel{>}{<}}(\omega)G^A(\omega),
\end{equation}
we have
\begin{eqnarray}
    J_l&=&\int\frac{d\omega}{2\pi }\sum_{kjj^\prime}\bar{V}^l_{kj}E^\prime_{k}\bar{V}^{l\ast}_{k j^\prime}\left[\left(G^R(\omega)\Sigma^>(\omega)G^A(\omega)\right)_{jj^\prime}\right.\nonumber\\
    &\times&g^{l<}_{k}(\omega)-\left.g^{l>}_{k}(\omega)\left(G^R(\omega)\Sigma^{{<}}(\omega)G^A(\omega)\right)_{jj^\prime}\right]. \label{eq:JL}
\end{eqnarray}

For the self-energy $\Sigma(\omega)$, we can calculate it from the second order of the S-matrix (in the Matsubara representation):
\begin{equation}
    G^{(2)}_{ij}(\omega)=\sum_{k k^{\prime}i^\prime j^{\prime}\alpha}G_{ii^{\prime}}(\omega)\bar{V}^{\alpha\ast}_{k,i^{\prime}}g^\alpha_{k}(\omega)\bar{V}^\alpha_{k j^{\prime}}G_{j^{\prime}j}(\omega).
\end{equation}
Here, $\alpha$ belongs to $l$ or $r$, denoting the left and right reservoirs,  and Eq.~\eqref{eq:V_Vconj} is used. From this equation, we get the self-energy as 
\begin{equation}
    \Sigma_{ij}(\omega)=\sum_{k\alpha}\bar{V}^{\alpha\ast}_{k i}g^\alpha_{k}(\omega)\bar{V}^\alpha_{k j},
\end{equation}
with 
\begin{eqnarray}
    g^{\alpha<}_{k}(\omega)&=&in_F(E_{k},T_\alpha)\delta(\omega-E_{k})2\pi,\\
    g^{\alpha>}_{k}(\omega)&=&-i[1-n_F(E_{k},T_\alpha)]\delta(\omega-E_{k})2\pi,\\
    g^{\alpha R/A}_{k}(\omega)&=&\mp i\delta(\omega-E_{k})\pi.
\end{eqnarray}
Here, $n_F(E_k,T_\alpha)$ is the Fermi distribution as defined in App.~\ref{app:thermal_current}. It is noted that the energy in the delta function is $E_{k}$ instead of $E^\prime_{k}$.  Further, define 
\begin{equation}
    \Gamma^\alpha_{ij}(\omega)=\sum_{k}\bar{V}^{\alpha\ast}_{ki}\bar{V}^\alpha_{k j}\delta(\omega-E_{k})2\pi, \label{eq:Gamma}
\end{equation}
 we have 
\begin{eqnarray}
 \Sigma^>_{ij}(\omega)&=&i\Gamma^l_{ij}(\omega)[n_F(\omega,T_l)-1]\nonumber\\
 &+&i\Gamma^r_{ij}(\omega)[n_F(\omega,T_r)-1],\\
    \Sigma^<_{ij}(\omega)&=&i\Gamma^l_{ij}(\omega)n_F(\omega,T_l)+i\Gamma^r_{ij}(\omega)n_F(\omega,T_r).\nonumber\\
\end{eqnarray}
Using $E^\prime_{k}=E_{k}/2$ and the $\delta(\omega-E_{k})$ in $g^{\alpha\stackrel{>}{<}}_{k}(\omega)$, we obtain
\begin{eqnarray}   
    J_l&=&\frac{1}{2}\int\frac{d\omega}{2\pi}\omega{\rm Tr}[G^R(\omega)\Gamma^r(\omega)G^A(\omega)\Gamma^l(\omega)]\nonumber\\
    &\times&[n_F(\omega,T_l)-n_F(\omega,T_r)].
\end{eqnarray}
This form is quite like the Landauer-B\"uttiker formula for normal metal, except for the factor $1/2$ and the "hole" part integration from the Nambu representation. It should be noted that although this formula is derived under the assumption that the system is in a steady state, an assumption that strictly does not hold for finite reservoirs, it remains valid in an approximate sense. If the reservoirs are sufficiently large, their internal states remain nearly unchanged over a certain period of time during which stable thermal currents are established. Within this time window, the Landauer-B\"uttiker formula remains a valuable tool for interpreting the results obtained from direct numerical simulations as shown in Fig.~\ref{Fig2}(d).

\section{\label{DeltaT} Robustness against reservoir temperature imbalance}
\begin{figure}[H]
    \centering
    \includegraphics[width=0.47\textwidth]{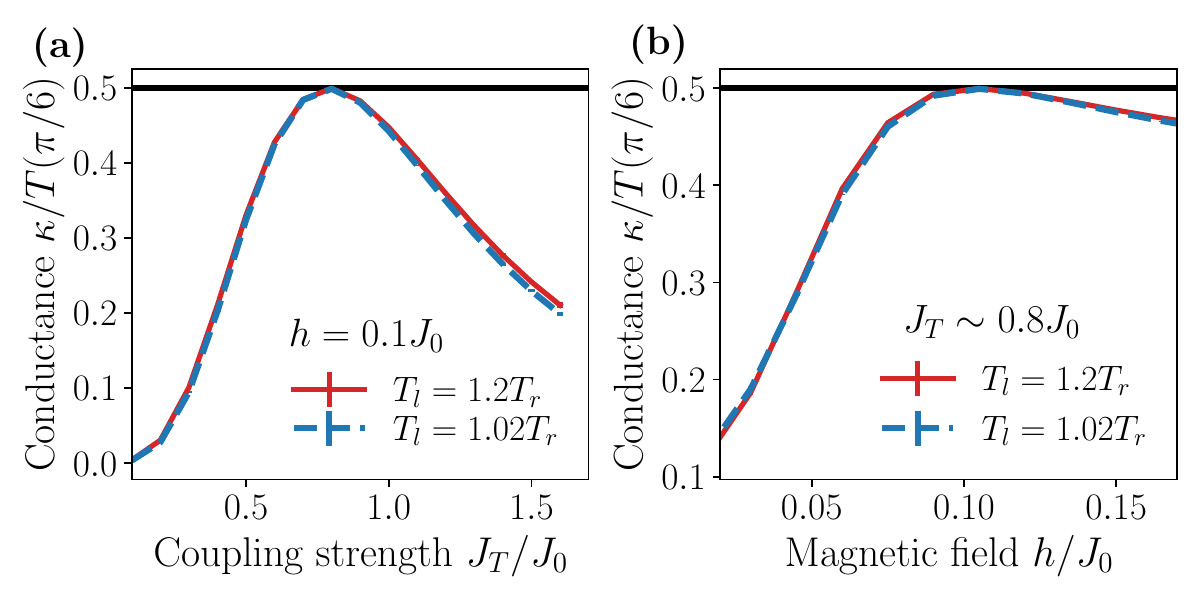}
    \caption{ \textbf{Robustness against reservoir temperature imbalance.} \textbf{(a)} and \textbf{(b)} Coupling strength $J_T/J_0$ and magnetic field strength $h/J_0$ dependence of thermal conductance at different temperature ratios $T_l/T_r$, respectively. Thermal conductance remains stable across different temperature ratios between the reservoirs, indicating that fine-tuning of reservoir temperatures is not essential for experimental realization. $T_l$ is always set to be $0.006J_0$ in the calculation. The Kitaev channel dimensions are $8_x\times16_y$, with each reservoir containing 200 sites.}
    \label{fig:deltaT}
\end{figure}

In our calculations, the reservoir temperatures are fixed with the ratio $T_l = 1.2 T_r$ for convenience. To demonstrate the effect of a smaller temperature difference, we also consider $T_l = 1.02 T_r$, as shown in Fig.~\ref{fig:deltaT}. The results show only minor differences between these cases when all other parameters are held constant. This indicates that precise fine-tuning of reservoir temperatures may not be essential in experimental implementations.

\section{\label{app:bulkgap}Bulk gap of the finite central chiral spin liquid channel}

\begin{figure}[]
\centering
\includegraphics[width=0.47\textwidth]{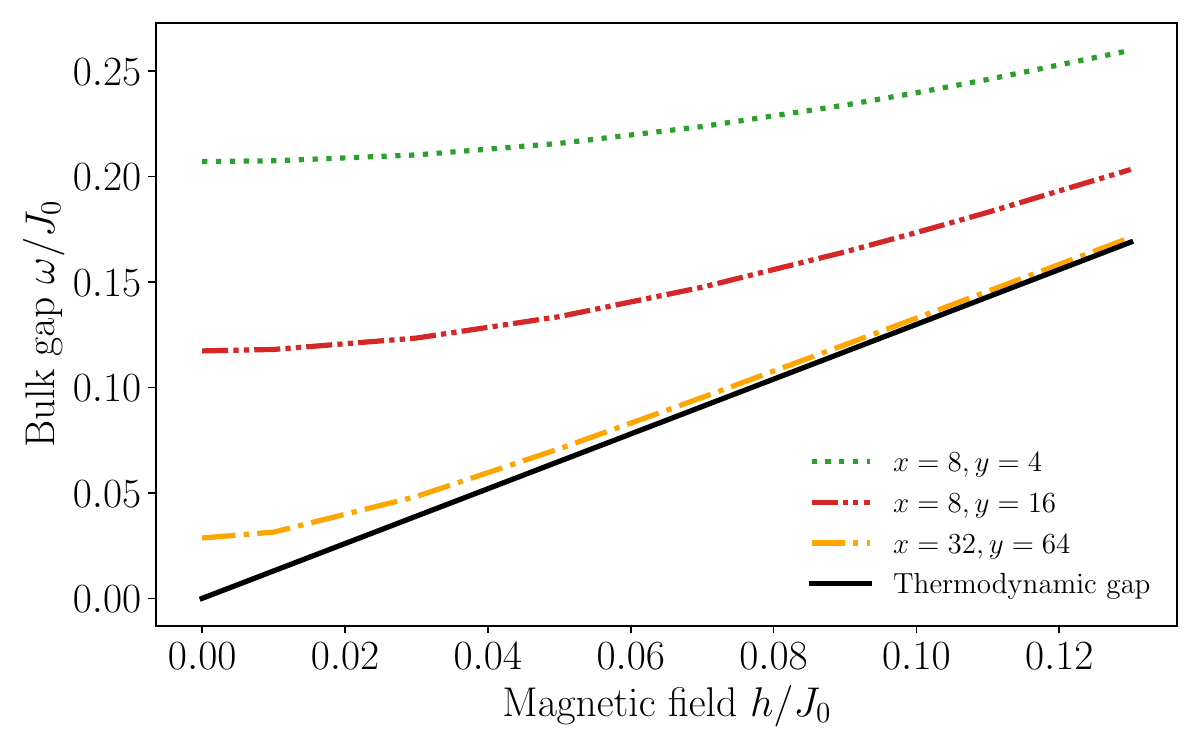}
\caption{\textbf{Magnetic field dependence of the bulk gap in the finite chiral spin liquid channel for different lengths along the $y$-direction.} The $x$ and $y$ denote the number of unit cells along the $x$ and $y$-direction, respectively. Due to finite-size effects, the bulk gap remains finite at zero field and increases with the magnetic field. As the system size increases, the bulk gap gradually approaches the thermodynamic limit (black solid line).}
\label{fig:bulkgap}
\end{figure}

In this section, we present the bulk energy gap of the finite central chiral spin liquid channel under various magnetic fields, as shown in Fig.~\ref{fig:bulkgap}. When open boundary conditions are applied along the $x$-direction, chiral edge states emerge within this bulk gap, and the corresponding transmission peaks observed in this energy region shown in Fig.~\ref{Fig4} originate from these edge states.

As a result of finite-size effects, the bulk gap remains nonzero even at zero magnetic field. However, increasing the number of sites tends to reduce the gap to the thermodynamic gap (black solid line in Fig.~\ref{fig:bulkgap}), and increasing the magnetic field leads to a monotonic enhancement of the bulk gap, as evident in Fig.~\ref{fig:bulkgap}.

\section{\label{app: few-sites} Few states model}
\begin{figure}[H]
    \centering
    \includegraphics[width=0.47\textwidth]{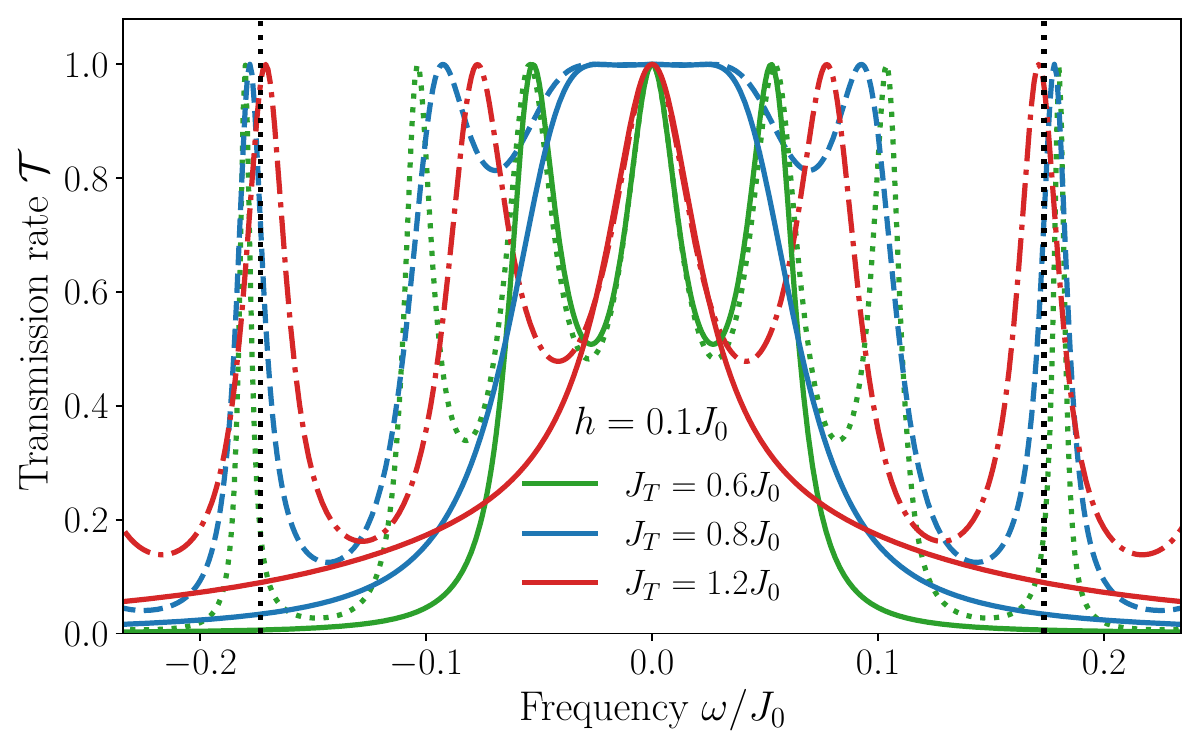}
    \caption{ \textbf{Comparison of transmission rates between the simplified and full models under varying coupling strengths at $h=0.1J_0$.} The green, blue, and red dashed curves indicate the transmission rate from the full model, while solid curves represent the simplified model, considering only the zero energy states and those nearest to them. Both models show agreement around $\omega=0$. The Kitaev channel dimensions are $8_x\times16_y$, with each reservoir containing 200 sites. The black vertical dotted line marks the location of the bulk energy gap $\Delta$ of the finite-size system.}
    \label{fig:few_full}
\end{figure}

As shown in Fig.~\ref{Fig4}, the thermal conductance is mainly contributed by the quasi-particle excitations near $\omega=0$. Thus,  to better understand the results, we can only keep the low energy state, i.e., the zero energy state and the states nearest to it. We can compare the results under this assumption with the one including all eigenstates, as shown in Fig.~\ref{fig:few_full}, where the transmission rates from both cases consist with each other quite well around $\omega=0$. 

Under this assumption, the corresponding Green function can be written as  
\begin{equation}
    \frac{1}{G^{R}(\omega)}=\left(\begin{array}{cccc}
   \omega & 0 & 0 & 0\\
    0 & \omega & 0 & 0\\
     0 & 0 & \omega & 0\\
      0 & 0 & 0 & \omega
    \end{array}\right)-\left(\begin{array}{cccc}
    -\epsilon & 0 & 0 & 0\\
    0 & 0 & 0 & 0\\
     0 & 0 & 0 & 0\\
      0 & 0 & 0 & \epsilon
    \end{array}\right)-i \frac{\Gamma^l+\Gamma^r}{2},
\end{equation}
and $G^A=G^{R\dagger}$. Here, $\epsilon$ is the energy of the nearest states.

As for the self-energy $\Gamma^{l/r}(\omega)$, it reflects both the reservoir density of states and the matrix elements of the coupling to individual system eigenstates, i.e., $\bar{V}^{\alpha}_{k i}$ as shown in Eq.~\eqref{eq:Gamma}. One can always choose the wave functions at the connection points to be real, which leads to a purely imaginary $\bar{V}^{\alpha}_{k i}$ and hence $\Gamma^\alpha_{ij}(\omega) = \Gamma^\alpha_{ji}(\omega)$. Furthermore, due to the particle-hole symmetry, we have $\bar{V}^{\alpha}_{k,j}=\bar{V}^{\alpha}_{k,-j}$, and hence, $\Gamma^\alpha_{ij}(\omega) = \Gamma^\alpha_{i,-j}(\omega)$. While an explicit analytic form of $\Gamma^{l/r}$ can in principle be derived, doing so requires detailed knowledge of the spatial structure of low-energy eigenstates near the system-reservoir boundary. Although such a calculation is straightforward numerically, it is analytically cumbersome and offers little additional physical insight. Importantly, since $\bar{V}^\alpha_{k i} \propto J_T$, the self-energy $\Gamma^\alpha_{ij}(\omega)$ scales as $J_T^2$. This monotonic relation implies that analyzing the dependence of thermal transport on $\Gamma^{l/r}$ effectively captures the qualitative influence of $J_T$. Therefore, for our simplified few-site model, we directly adopt a phenomenological form of $\Gamma^{l/r}$ consistent with its symmetry properties:
\begin{eqnarray}
    \Gamma^l&=&\left(\begin{array}{cccc}
    A_1 & A_2 & A_2 & A_1\\
    A_2 & A_0 & A_0 & A_2\\
    A_2 & A_0 & A_0 & A_2\\
    A_1 & A_2 & A_2 & A_1
    \end{array}
    \right)\label{GammaL}\\
    \Gamma^r&=&\left(\begin{array}{cccc}
    A_1 & -A_2 & -A_2 & A_1\\
    -A_2 & A_0 & A_0 & -A_2\\
    -A_2 & A_0 & A_0 & -A_2\\
    A_1 & -A_2 & -A_2 & A_1
    \end{array}
    \right).\label{GammaR}
\end{eqnarray}
Here, for simplicity, we omit the frequency dependence of $\Gamma^{l/r}$, as our analysis is restricted to the small $\omega$ regime. It is noted that the difference between $\Gamma^l$ and $\Gamma^r$ comes from the different phases for the two points connecting the reservoirs (the second reservoir is at the opposite point of the first reservoir, with the corresponding phase being $\pi$), and the transmission rate can be obtained as 
\begin{widetext}
\begin{equation}
    \mathcal{T}=\frac{4(-8A_0A_1A_2^2\omega^2+A_1^2\omega^4+2A_2^2\omega^2(\epsilon^2-\omega^2)+A_0^2(\epsilon^4+8A_1^2\omega^2-2\epsilon^2\omega^2+\omega^4))}{(4A^2_0+\omega^2)(-\epsilon^2+\omega(-2i A_1+\omega))(-\epsilon^2+\omega(2iA_1+\omega))}.
\end{equation}
\end{widetext}

With these definitions, one can obtain a quite complex transmission rate from the formula of transmission rate Eq.~\eqref{eq:trans}. However, considering that the relevant $\omega$ is very small, we can expand this equation up to the second order of $\omega$ as 
\begin{equation}
    \mathcal{T}\sim 1-\frac{-16(A_0A_1-A_2^2)^2+(4A_2^2-\epsilon)^2}{4A^2_0\epsilon^4}\omega^2+{\rm O}(\omega^4).
\end{equation}
Since half-integer thermal conductance requires an almost flat transmission near $\omega=0$, the coefficient of the $\omega^2$ term should vanish. Neglecting the $k$ dependence of real $\bar{V}^{\alpha}_{k i}$ in the low-energy limit, Eq.~\eqref{eq:Gamma} yields $A_2^2 \sim A_0 A_1$. Under this approximation, the vanishing condition for the $\omega^2$ coefficient becomes $\epsilon = 2\sqrt{A_0 A_1}$. Moreover, under this condition, the transmission rate becomes 
\begin{equation}
    \mathcal{T}\sim 1-\frac{\omega^6}{\epsilon^4 4A_0^2}+{\rm O}(\omega^8).
\end{equation}
If we further assume that the coupling strengths between the reservoirs and the two lowest-energy eigenstates of the Kitaev channel are approximately equal, i.e., their boundary wavefunction distributions are nearly identical, one can take $A_0 = A_1$, leading to $4A_0^2 = \epsilon^2$. The transmission rate then simplifies to
\begin{equation}
    \mathcal{T} \sim 1 - \frac{\omega^6}{\epsilon^6} + \mathcal{O}(\omega^8).
\end{equation}
This result shows that the transmission rate is almost unity around $\omega = 0$ until $\omega$ is close to $\epsilon$, which is just the plateau observed in Fig.~\ref{Fig4}. This also tells us that the width of the plateau scales with the energy spacing $\epsilon$ between the two lowest eigenstates.

On the other hand, considering that $A_0$, $A_1$, and $A_2$ are proportional to $J_T^2$, there would exist a optimal $J_T$ fulfills this relationship $\epsilon = 2\sqrt{A_0 A_1}$. However, if the magnetic field further increases, $\epsilon$ also increases, which breaks the best $J_T\sim\epsilon$ pair. Nevertheless, the coefficient of the $\omega^2$ term is still the smaller the better. Thus, we can conclude that the larger the $\epsilon$, the better the results. This also explains the observation exhibited in Fig.~\ref{Fig4} that a smaller magnetic field corresponds to a narrower transmission plateau.

\begin{figure}[]
    \centering
    \includegraphics[width=0.47\textwidth]{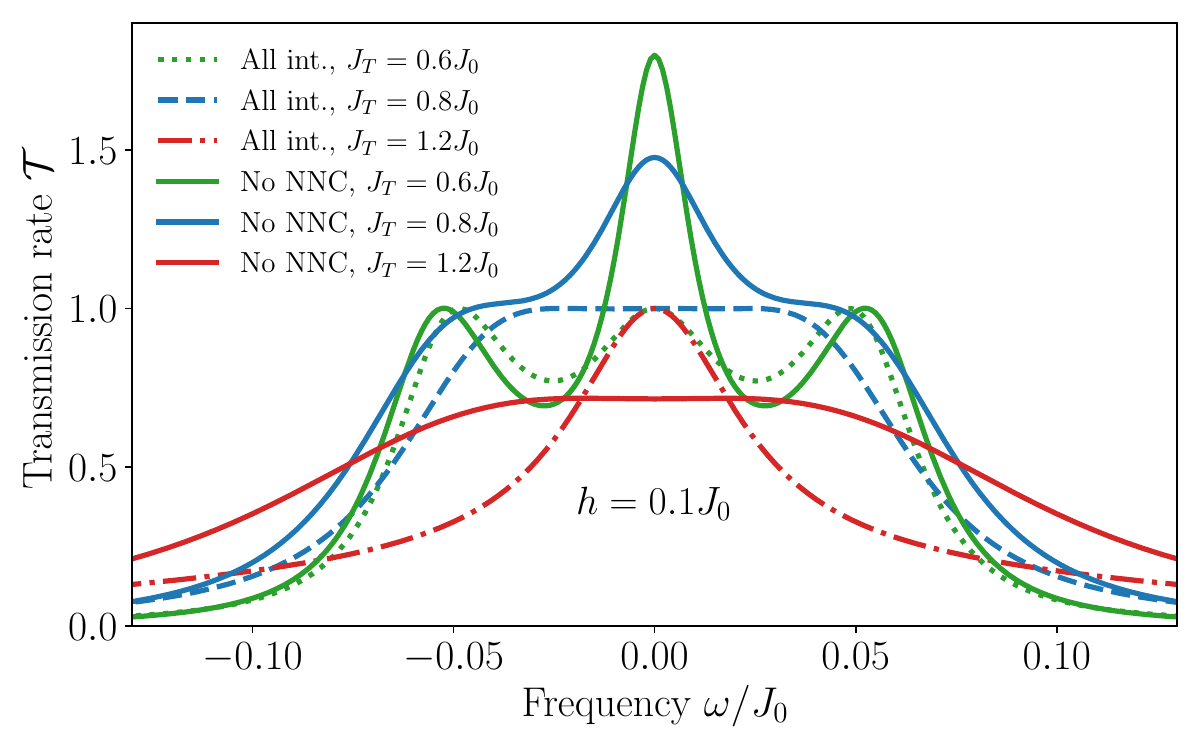}
    \caption{ \textbf{Influence of the number-non-conserving (NNC) terms from the Majorana nature of the system.} Dashed curves show the transmission with all interactions included, while solid curves exclude the number-non-conserving terms (NNC, as marked in the figure). The absence of NNC terms significantly alters the transmission near $\omega=0$, deviating it from unity and thereby obstructing the attainment of the half-quantized thermal conductance. The other parameters are the same as in Fig.~\ref{fig:few_full}. }
   \label{fig:Maj_Norm}
\end{figure}

Building upon the discussion of the simplified model, we explore the origin of the half-quantized thermal conductance under the Landauer-B\"uttiker framework. Our findings demonstrate that the observed plateaus are a direct consequence of the Majorana fermion nature of the system. Specifically, due to the Majorana form [Eqs.~\eqref{eq:HR} and (\ref{MajH})] of the system, the coupling between reservoirs and the transport channel in the Nambu representation has number-non-conserving terms [e.g., the $\beta^\dagger_k\gamma^\dagger_j$ and their hermitian terms with $k>0$ and $j>0$ in Eq.~\eqref{eq:Vkpm}], which leads to off-diagonal terms in the $\Gamma$ matrix [Eqs.~(\ref{GammaL}) and (\ref{GammaR})]. These terms are essential for the observed results. Without them, the transport reduces to that of conventional fermionic systems, losing the distinctive Majorana signatures. To illustrate the significance of these off-diagonal terms, we modify the simplified model by removing them and presenting the corresponding results in Fig.~\ref{fig:Maj_Norm}. Different from the results with these terms (dashed curves), eliminating these terms (solid curves) drives the transmission near $\omega=0$ significantly away from one and consequently prevents the thermal conductance from achieving the half-quantized value. This contrast highlights the fundamental role of Majorana fermions in sustaining the observed transport characteristics.

\section{\label{app:LB_half}Influence of the Majorana fermions}
In the previous section, we showed numerically that the zero-energy Majorana fermions play an important role on the appearance of the half-quantized thermal conductance. In this section, we try to explain this theoretically by using a simplified toy model: only consider the transmission through the two Majorana states. In this case, we can always write the $\Gamma^{L/R}$ matrix as 
\begin{equation}
    \Gamma^{L/R}(\omega)=\left(\begin{array}{cc}
        \Gamma^{L/R}_{1,1}(\omega) & \Gamma^{L/R}_{1,-1}(\omega)\\ 
        \Gamma^{L/R}_{-1,1}(\omega) & \Gamma^{L/R}_{-1,-1}(\omega) 
    \end{array}\right).
\end{equation}
According to Eqs.~\eqref{eq:V_Vconj} and~\eqref{eq:Gamma}, we have 
\begin{equation}
    \Gamma^\alpha_{ij}(\omega)=\Gamma^{\alpha\ast}_{ji}(\omega)=\Gamma^{\alpha}_{-j,-i}(-\omega).
\end{equation}
Therefore, we have $\Gamma^\alpha_{1,1}(\omega\rightarrow0)=\Gamma^\alpha_{-1,-1}(\omega\rightarrow0)$ being real, and $\Gamma^\alpha_{1,-1}(\omega\rightarrow0)=\Gamma^{\alpha\ast}_{-1,1}(\omega\rightarrow0)$. Then, for the transmission through the Majorana fermions, the Green function can be written as:
\begin{equation}
    G^{R/A}(\omega)=\frac{1}{\left(\begin{array}{cc}
        \omega & 0\\ 
        0 & \omega
    \end{array}\right)\pm i\frac{\Gamma^L(\omega)+\Gamma^R(\omega)}{2}}.
\end{equation}
With these parameters and $\Gamma^L=\Gamma^R=\Gamma$ in our model, the transmission rate around $\omega=0$ can be obtained as 
\begin{eqnarray}
&&{\rm Tr}[G^R(\omega)\Gamma^R(\omega)G^A(\omega)\Gamma^A(\omega)]\nonumber \\
&=&\frac{2 \left(\Gamma_{11}^4+\Gamma_{11}^2 \left(\omega^2-2 |\Gamma_{1,-1}|^2\right)+|\Gamma_{1,-1}|^2 \left(|\Gamma_{1,-1}|^2+\omega^2\right)\right)}{\Gamma_{11}^4+\Gamma_{11}^2 \left(2 \omega ^2-2 |\Gamma_{1,-1}|^2\right)+\left(|\Gamma_{1,-1}|^2+\omega^2\right)^2}. \nonumber\\
\end{eqnarray}
For normal cases, i.e., $\Gamma_{11}\ne|\Gamma_{1,-1}|$, we can get ${\rm Tr}[G^R(0)\Gamma^R(0)G^A(0)\Gamma^A(0)]=2$ and then we have
\begin{eqnarray}
    J_L&\approx&\int \frac{d\epsilon}{2\pi}\epsilon [f^L(\epsilon)-f^R(\epsilon)] \nonumber \\ &\approx&\int \frac{d\epsilon}{2\pi}\epsilon \partial_Tf(\epsilon)\nabla T=\frac{\pi T}{6}\nabla T.
\end{eqnarray}
From the definition of thermal conductance $J_Q=-\kappa \nabla T$, we can obtain that $\kappa/T={\pi}/{6}$, which is quantized instead of half-quantized. 

Nevertheless, for our system with $|\Gamma_{1,1}|=|\Gamma_{1,-1}|=|\Gamma_{-1,1}|=|\Gamma_{-1,-1}|$ due to the particle-hole symmetry [i.e., $|V^\alpha_{k,j}|=|V^\alpha_{k,-j}|$ in Eq.~\eqref{eq:Gamma}], there are exceptions. The Green function $G^{R/A}(\omega=0)$ does not exist because $\Gamma^{L/R}(0)$ has zero eigenvalues and its inverse matrix does not exist. However, for $\omega\ne0$, the inverse matrix still exists, and we can find that in this case $\lim_{\omega\rightarrow 0}{\rm Tr}[G^R(\omega)\Gamma^R(\omega)G^A(\omega)\Gamma^A(\omega)]=\lim_{\omega\rightarrow 0}4\Gamma^2_{11}/(4\Gamma^2_{11}+\omega)=1$ instead of 2, which corresponds to half the quantized thermal conductance. 

\section{\label{app:Chern} Evaluating Topology in Small Systems with Twisted Boundary Conditions}
In this section, we discuss how twisted boundary conditions can be used to compute the topological invariant in finite-size systems \cite{niu1985quantized, nathan2024relating}. For a finite system with Hamiltonian $H$, we introduce twisted boundary conditions:
\begin{eqnarray}
    H(x+L_x,y)&=&e^{i\theta_x} H(x,y),\\
    H(x,y+L_y)&=&e^{i\theta_y} H(x,y).
\end{eqnarray}
where $L_x$ and $L_y$ are the system lengths in the $x$- and $y$-directions, respectively. The Hamiltonian then becomes a function of the twist angles, denoted by $H(\theta_x,\theta_y)$, with the corresponding eigenstates determined by
\begin{equation}
H(\theta_x,\theta_y)|\phi(\theta_x,\theta_y)\rangle=E(\theta_x,\theta_y)|\phi(\theta_x,\theta_y)\rangle.
\end{equation}
The parameters $\theta_x$, $\theta_y\in[0,2\pi)$ define a closed two-dimensional space. As long as the ground state $\phi_0$ is non-degenerate and smooth in the $(\theta_x,\theta_y)$ space, the Berry curvature can be defined as
\begin{equation}
    \mathcal{F}(\theta_x,\theta_y)=i\left(\left\langle\frac{\partial \phi_0}{\partial \theta_x}\middle|\frac{\partial\phi_0}{\partial\theta_y}\right\rangle-\left\langle\frac{\partial \phi_0}{\partial \theta_y}\middle|\frac{\partial\phi_0}{\partial\theta_x}\right\rangle\right).
\end{equation}
If the ground-state band $E_0(\theta_x,\theta_y)$ is gapped from the excitation state band, the many-body Chern number of this finite-size system can be defined as 
\begin{equation}
    \nu=\frac{1}{2\pi}\int^{2\pi}_0 d\theta_x\int^{2\pi}_0d\theta_y \mathcal{F}(\theta_x,\theta_y).
\end{equation}
This twisted boundary condition approach does not rely on the specific system size and therefore enables accurate evaluation of the topological invariant even in small finite systems.

\section{\label{LargerSystem} Thermal conductance in large system}
\begin{figure}[]
    \centering
    \includegraphics[width=0.47\textwidth]{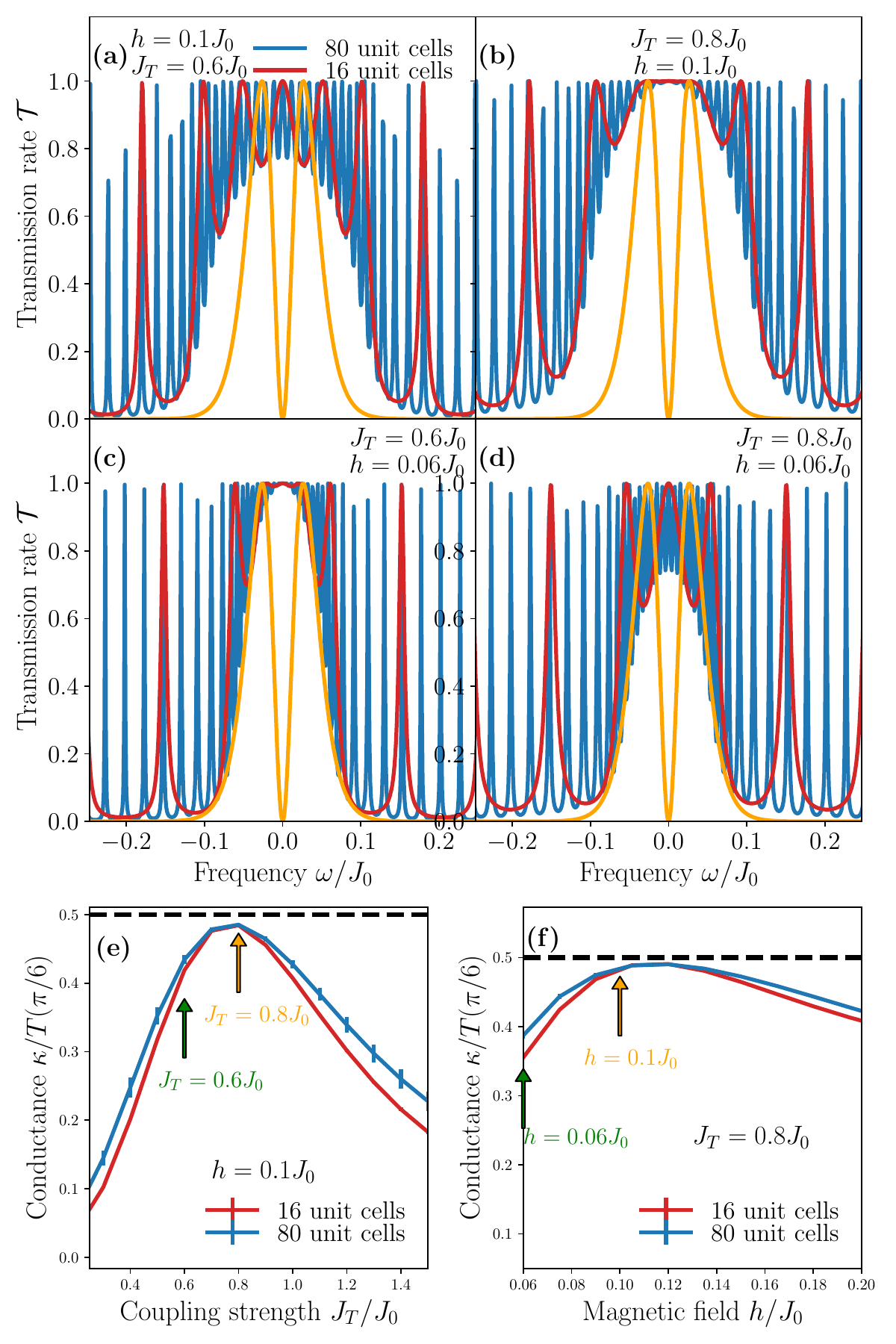}
    \caption{ \textbf{Influence of larger $L_y$} (a)-(d) Transmission spectrum for $L_y=16$ (red curve) and $L_y=80$ (blue curve) under different $h$ and $J_T$. In the gap region, dense peaks corresponding to the chiral edge states appear when $L_y$ is large. The orange curve is the normalized energy distribution difference $|\omega[n_F(\omega,T_l)-n_F(\omega,T_r)]|/C$, where $n_F$ is the Fermi distribution and $C$ is a normal factor for visualization. 
    (e) and (f) Thermal conductance under varying $J_T$ and $h$, respectively. Increasing $L_y$ helps the thermal conductance approach the half-quantized value (the black dashed line).}
   \label{fig:Larger}
\end{figure}

Our investigation mainly focuses on systems with fewer lattice sites along the $y$-direction, where the half-quantized thermal conductance is primarily achieved through the finite-size effect induced transmission peaks/plateau at the Majorana zero modes. While these small systems like 16 unit cells along the $y$ direction, i.e., $L_y=16$, can reach half-quantized conductance, they require extremely low temperatures and differ from the conventional understanding that associates half-quantized conductance with chiral edge states. To further investigate this issue, we examine systems with larger $y$-direction lattice sizes, as shown in Fig.~\ref{fig:Larger}.

Panels (a) to (d) of Fig.~\ref{fig:Larger} compare the transmission spectra at different magnetic fields $h$ and coupling strength $J_T$ values for two cases: one with a smaller lattice size ($L_y = 16$, red curve) and one with a larger lattice size ($L_y = 80$, blue curve). The transmission peaks correspond to the eigenstates, with the transmission valleys between adjacent energy levels. For higher temperatures, the excitations are more broadly distributed, as demonstrated by the energy distribution difference used in Eq.~\eqref{eq:Landauer} $|\omega[n_F(\omega,T_l)-n_F(\omega,T_r)]|/C$ shown in the orange curve. Here, $C$ is a normal factor for visualization in the figure. Notably, a significant portion of the excitations lies within the transmission valleys. Since achieving half-quantized thermal conductance requires unit transmission at the relevant excitation energies, the system cannot achieve half-quantized conductance due to these valleys of the transmission spectrum.

However, when $J_T$ approaches the optimal value [Fig.~\ref{fig:Larger}(b) and (c)], these valleys can be suppressed, leading to an increase in thermal conductance. As the lattice size increases, more transmission peaks emerge, filling in the valleys as shown in Fig.~\ref{fig:Larger}(a) and (d), which enhance the conductance. This trend is confirmed in Fig.~\ref{fig:Larger}(e) and (f), where the thermal conductance for $L_y = 16$ (red curve) and $L_y = 80$ (blue curve) is plotted against $J_T$ and $h$, respectively. The results show a clear increase in thermal conductance as the lattice size increases. We infer that, as the lattice size approaches infinity, the finite-size effects vanish, and the system transitions back to the traditional understanding of half-quantized thermal conductance coming from the chiral edge states.

\bibliography{ref}

\end{document}